\documentclass[12pt,superscriptaddress,notitlepage,nofootinbib]{revtex4-1}
\usepackage[utf8x]{inputenc}
\usepackage[T1]{fontenc}
\usepackage{graphicx}
\usepackage{color}
\usepackage{mathtools}
\usepackage{amssymb}
\usepackage{amsthm}

\begin{document}
\title{A trace formula for metric graphs
with piecewise constant potentials and multi-mode graphs}
\author{Sven~Gnutzmann}
\email{sven.gnutzmann@nottingham.ac.uk}
\affiliation{School of Mathematical Sciences, University of Nottingham, University Park, Nottingham NG7 2RD, UK}
\author{Uzy~Smilansky}
\email{uzy.smilansky@weizmann.ac.il}
\affiliation{Department of Physics of Complex Systems, Weizmann Institute of Science, Rehovot 76100, Israel}
\affiliation{Department of Mathematical Sciences, University of Bath, Bath BA2 7AY, UK}
\begin{abstract}
  We generalize the scattering approach to quantum graphs to
  quantum graphs with with piecewise
  constant potentials and multiple excitation modes.
  The free single-mode case is well-known and leads to
  the trace formulas of Roth \cite{roth},
  Kottos and Smilansky \cite{KS_traceformula}.
  By introducing an effective reduced scattering picture we are
  able to introduce new exact trace formulas in the more general setting.
  The latter are derived and discussed in details with some
  numerical examples for illustration.\\
  Our generalization is motivated by both experimental applications
  and fundamental theoretical considerations. The free single-mode quantum
  graphs are an extreme idealization of reality that, due to the simplicity
  of the model allows to understand a large number of generic or universal
  phenomena. We lift some of this idealization by considering
  the influence of evanescent modes that only open above threshold energies.
  How to do this theoretically in a closed model in general is a challenging
  question of fundamental
  theoretical interest and we achieve this here for quantum graphs.
\end{abstract}

\maketitle

\section*{\textit{Dedication}}

\textit{ 
  This article is dedicated to the memories of Fritz Haake and
  Petr Braun.}
  
\textit{Fritz has been an outstanding personality.
  For me (Sven) he was my teacher, Doktorvater,
  mentor and friend. A role model in science and in life.
  For me (Uzy) he was a friend, a
  colleague and a companion in many events and inspiring discussions.\\
  \hspace*{1cm} \hfill Sven Gnutzmann and Uzy Smilansky}

\section{Introduction}

Metric graphs with a self-adjoint wave operator, known as quantum
graphs, turned out to be a paradigmatic model in the physics of
complex wave systems (quantum chaos \cite{KS_traceformula, review})
and in mathematical spectral theory \cite{grisha_book}.  At the same
time the model was applied to wave properties of actual physical
networks such as e.g., optical fibers, microwave cables or waveguides
\cite{exp_hul,exp_allgaier,exp_rehemanjiang,exp_dietz,exp_fu,johnson,exp_lawniczak}.
For most of these applications the quantum graph model suffices, in
spite of its being a drastic idealization of the complete physical
system: It is limited to complex-valued scalar wave functions that
propagate freely along the edges and their scattering in the junctions
(vertices) are provided by appropriate boundary conditions. In spite
of this idealization the quantum graph models grasp the essential
characteristics of the systems under study, and has the attractive
feature that it is simple in structure, and enables numerical
simulations of a scale which is prohibitive for more `realistic'
models.  One of the most prominent successes of quantum graphs is in
providing a rich, versatile and non-trivial spectral theory. The main
tool in this direction was the use of a scattering approach
\cite{KS_traceformula} to derive a secular function whose zeros
coincide with the wave operator spectrum. Moreover, this secular
function provides the basis for deriving a trace formula
\cite{KS_traceformula, review} for the spectral counting function
\(N(E)=\#\{E_n\in \sigma (G) : E_n<E\}\), where \(\sigma(G)\) is the
spectrum of the wave operator on the metric graph \(G\)
arranged in a non-decreasing order. This trace formula describes the
spectral counting function as a sum of two terms: \textit{i.}
A smooth one which
accounts for the mean increase of \(N(E)\), known as the Weyl-term.
For a graph of total length \(L_{\mathrm{tot}}\)
reads
\( \overline{N} (E) = \sqrt{E} L_{\mathrm{tot}}/\pi +\mathcal{O}(1)\).
\textit{ii.} An oscillatory term \(N_{\mathrm{osc}}(E)\) which can be written as
a sum of amplitudes computed for all the periodic orbits in the
graph. Each amplitude here is an oscillating function of the wave number
\(k=\sqrt{E}\). 
  
The purpose of the present work is to generalize the simple quantum
graph model so as to enable the study of realistic networks and at
the same time to retain as much as possible the simplicity of the
quantum graph model. The main focus will be on providing a
scattering approach and an extended trace formula which surmounts
the conceptual and technical difficulties posed by the physical
problem.

A realistic network is composed of waveguides and junctions where
several waveguides are connected. The waveguides are assumed to be
straight, with a constant transversal cross
section (\cite{exner_book, olaf_book} are recommended for a
detailed study of waveguides and networks).  The longitudinal and
transversal degrees of freedom are separated and therefore, the
wave functions in the edges are super-positions of product
functions
\begin{equation}
  \Psi_E(x,y) = \sum_{n=0}^{\infty} \left[ a^{+}_n f_n (y)
    \exp\left(ik_n(E)x\right) +a^{-}_n f_n (y) \exp\left(-ik_n(E)x\right)\right]
  \label{one}
\end{equation}
where \(E\) is the total energy, \(x\) stands for the longitudinal
degree of freedom, and \(y\) stands for the collections of
transversal degrees of freedom. \(f_n(y)\) are the transversal
mode eigenfunctions corresponding to energies \(\epsilon_n\).
\(k_n (E)=\sqrt{E-\epsilon_n}\) is the wave number in the
longitudinal direction if $E>\epsilon_n$ and the rate of
exponential decay or increase if \(E<\epsilon_n\).  The complex
coefficients \( a^{\pm}_n\) are the amplitudes of the waves
counter propagating (or decaying/increasing) in the longitudinal
direction. The main complication in this multi-mode approach is due
to the fact that the number of propagating modes (for which
\(E >\epsilon _n\)) increases by one whenever \(E\) crosses the
``threshold'' \(\epsilon_n\). As a result, the analytic structure of
the wave functions is complicated in a drastic way. The quantum
graph model is constructed by reducing the transversal size to
zero, thus pushing \(\epsilon_1\) far away so that the range
\(\epsilon_0<E <\epsilon_1\) becomes large, and one can ignore all
but the ground state mode. The approximation taken here is to
truncate the infinite sum in \eqref{one} at a finite number of
modes \(N_m\). Thus, one has to face the treatment of the
singularities at threshold -- a challenge that is addressed in the
present paper. Note that the transversal dynamics may differ
between different waveguides in the network so that the mode
spectra can induce a rich plethora of thresholds and
singularities.
 
The other elements in the network are the junctions. They can be
considered as cavities which couple to the waveguides in a known
way. The full computation of the spectrum for a general network
involves the spectrum in the entire enclosed volume which is
practical only for very simple networks. The engineering approach
is to measure the scattering matrix of the relevant junctions, and
they are used in the further computation.  This is also quite
cumbersome. The reduction of a network to a quantum graph solves
this problem by reducing the size of the junction cavity together
with the reduction of the transversal size \cite{olaf_book}, which
result in deriving effective boundary conditions at the junctions
(now vertices). The approximation chosen here is to generalize the
boundary conditions in an appropriate way -- which ensures the
conservation of flux in the systems or expressing this in more
mathematical terms ensures the self adjoint nature of the wave
operator.

The model which results from the two approximations – truncation
of the number of modes, and replacing the junction by boundary
conditions at the vertices, is the multi-mode graphs which appears
in the title of the present article.  This model will be denoted
by MM (for Multi-Mode) in the sequel.

As it stands, the MM model can be further reduced to the solution
of quantum graphs in which the edges \(e\) are endowed with
constant potentials \(V(e)\) . Then, an edge \(e\) allows free
propagation if \(E>V(e)\) and becomes evanescent otherwise. This
model (to be denoted by PCP for Piecewise Constant Potentials)
needs to include the proper treatment of thresholds as the MM
model.  The PCP model retains however only a single degree of
freedom as is the case for a quantum graph.  The interaction
between waves comes to play by taking advantage of the freedom in
the vertex boundary conditions in this quantum graphs.  Thus, for
any MM quantum graph, one can construct a PCP quantum graph which
has the same spectrum as the MM graph and equivalent
eigenfunctions. This is done by replacing each edge in the MM
model by \(N_m(e)\) parallel edges with potentials
\(V_m (e) = \epsilon_m (e) \). Due to this property, we shall
focus on the solutions of the PCP models, and indicate how to
connect it to the desired MM model using the wealth of allowed
boundary conditions.
 
The study of the role of thresholds and evanescent modes was
carried out in the literature for a few systems 
\cite{weidenmueller,schanz,rouvinez}. While we are not aware of such a
study for
quantum graphs, our derivation of a trace formula is based on an
approach by Brewer, Creagh and Tanner \cite{brewer} who
considered analogous generalizations in the context of an
elastic network of plates which has many features in common with
quantum graphs.

This manuscript is organized as follows: In Section
\ref{sec:interval} we discuss an interval with a potential step as
an introductory example that contains most of the essential
ingredients of the more general theory in a simple setting. In
Section \ref{sec:graph_operator} we define Schr\"odinger operators
on PCP quantum graphs.
We then introduce the Schr\"odinger operator for
MM  quantum graphs, and show that MM graphs 
can equivalently be described as PCP graphs
on an enlarged metric graph.  In
Section \ref{sec:scattering} we develop the scattering approach
for PCP (and hence MM) graphs
which generally leads to non-unitary scattering matrices and
quantum maps due to the presence of evanescent modes. Unitarity is
replaced by a different set of symmetries that we derive from
first principles.  In Section \ref{sec:trace} we derive a trace
formula for the spectral counting function for PCP graphs using the scattering
approach and illustrate our results with some numerical examples.
We conclude the paper in Section \ref{sec:conclusion} with an
outlook on experimental and theoretical applications and open
questions.

\section{Introductory example: an interval with a potential step}
\label{sec:interval}

Before going into the full-blown theory we discuss a simple example that
already
contains some of the main ideas: a quantum particle confined to an interval
with a potential step
described by the stationary Schr\"odinger equation
\begin{equation}
  - \phi''(x) + V(x) \phi(x) = E \phi(x)
\end{equation}
on the interval \( x\in [0,L] \)
of length \(L>0\).
Here, \(V(x)\) is a piecewise constant potential with one
potential step. Writing \(L=L_1+L_2\) with \(L_1>0\) and \(L_2>0\)
we write this potential step as
\begin{equation}
 V(x) =
 \begin{cases}
   0 & \text{for \( x\in [0,L_1)\),}\\
   V & \text{for \( x\in (L_1,L]\)}
 \end{cases}
\end{equation}
with \(V>0\). See \cite{weidenmueller} where, among other,
the open variant of
this model was discussed from a pure scattering point of view.
At the ends of the interval we
require Dirichlet conditions \( \phi(0)=\phi(L)=0 \) and
at \(x=L_1\) we require that the wave function and its derivative are
continuous, \( \phi(L_1^-) = \phi(L_1^+) \) and \(\phi'(L_1^-)= \phi'(L_1^+) \)
where the notation \(L_1^\pm\) indicates the limits from above and below.
Our conditions imply that the stationary Schr\"odinger equation describes a
self-adjoint eigenvalue problem with a purely positive spectrum. We thus assume
\(E>0\) in the following.
One may view this setting as a quantum star graph with two edges of lengths
\(L_1\) and \(L_2\) and edgewise constant potentials.
Accordingly we will refer to the subintervals \( [0,L_1)\)
and \( (L_1,L]\) as edges and the position \(x=L_1\) as the central vertex.
Let us introduce the wavenumbers 
\begin{equation}
  k_1= \sqrt{E}= k \quad \text{and} \quad k_2=\sqrt{E-V} 
\end{equation}
and note that \(k_2\) is real only if \(E\ge V\) while it is
purely imaginary for small energies  \(E < V\). In the latter case we
choose to have a positive imaginary part \( k_2 = i |k_2| \).
We may construct solutions starting from a
superposition of plane waves with unit
fluxes
\begin{equation}
  \phi(x)
  =
  \begin{cases}
    \frac{1}{\sqrt{k_1}}\left(b_1^{\textrm{in}}  e^{ik_1 (x-L_1)} +b_1^{\textrm{out}}
      e^{-ik_1 (x-L_1)} \right) & \text{for \( x\in [0,L_1)\),}\\
    
    \frac{1}{\sqrt{k_2}}\left(b_2^{\textrm{in}}  e^{-ik_2 (x-L_1)} +b_2^{\textrm{out}}
      e^{ik_2 (x-L_1)} \right) 
    & \text{for \( x\in (L_1,L]\)}
  \end{cases}
\end{equation}
where \( b_{1/2}^{\mathrm{in/out}} \) are the (complex)
amplitudes of in-/outgoing plane waves at the potential step \(x=L_1\).
Note that for \(E<V\) on the edge \( (L_1,L]\) one has real exponents that describe
exponential decay or increase -- in this case we have implicitly
defined the direction of
propagation as the direction of decay.
The conditions at the central vertex \(x=L_1\) may be
now be written as
\begin{equation}
  \begin{pmatrix}
    b_1^{\textrm{out}}\\
    b_2^{\textrm{out}}
  \end{pmatrix} = \sigma (E)
  \begin{pmatrix}
    b_1^{\textrm{in}}\\
    b_2^{\textrm{in}}
  \end{pmatrix}
\end{equation}
with the energy dependent central vertex scattering matrix
\begin{equation}
  \sigma(E) =
  \begin{pmatrix}
    \frac{k_1 -k_2}{k_1+ k_2} & \frac{2\sqrt{k_1 k_2}}{k_1+k_2}\\
    \frac{2\sqrt{k_1 k_2}}{k_1 + k_2} & - \frac{k_1 -k_2}{k_1+ k_2}
  \end{pmatrix} .
  \label{eq:example_central_scattering}
\end{equation}
Furthermore the Dirichlet conditions at the outer vertices imply
\begin{equation}
  \begin{pmatrix}
    b_1^{\textrm{in}}\\
    b_2^{\textrm{in}}
  \end{pmatrix} = \tau (E)
  \begin{pmatrix}
    b_1^{\textrm{out}}\\
    b_2^{\textrm{out}}
  \end{pmatrix}
\end{equation}
where the diagonal matrix
\begin{equation}
  \tau(E) =
  \begin{pmatrix}
    - e^{2 i k_1 L_1} & 0\\
    0 & - e^{2i k_2 L_2}
  \end{pmatrix} 
\end{equation}
contains the phases that are acquired by going along the edge,
being reflected and then coming back to the center.
It is straight forward to check that \(\sigma(E)\) and \( \tau(E) \)
are unitary for
\( E> V \).

Altogether
the energy \(E \neq V\) is an eigenvalue if and only if the
consistency condition
\begin{equation}
  \begin{pmatrix}
    b_1^{\textrm{in}}\\
    b_2^{\textrm{in}}
  \end{pmatrix} = U (E)
  \begin{pmatrix}
    b_1^{\textrm{in}}\\
    b_2^{\textrm{in}}
  \end{pmatrix}
  \label{quantisation_condition}
\end{equation}
with the quantum map
\begin{equation}
  U(E) =\tau(E) \sigma(E) =
  \begin{pmatrix}
    -\frac{k_1 -k_2}{k_1+ k_2}e^{2i L_1 k_1} &
    -\frac{2\sqrt{k_1 k_2}}{k_1+k_2}e^{2i L_1 k_1}\\
    -\frac{2\sqrt{k_1 k_2}}{k_1 + k_2}e^{2i L_2 k_2} &
    \frac{k_1 -k_2}{k_1+ k_2}e^{2i L_2 k_2}
  \end{pmatrix}
  =
  \begin{pmatrix}
    U_{11} & U_{12}\\
    U_{21} & U_{22}
  \end{pmatrix} 
\end{equation}
is satisfied in a non-trivial way.
This is equivalent to the condition that the secular function
\(\xi(E)\) vanishes, where 
\begin{equation}
  \xi(E)= \det\left(\mathbb{I}-U(E)\right)\ .
  \label{eq:xi_at}
\end{equation}
The correspondence between energy eigenvalues and zeros of the
secular function is one-to-one for all real energies apart from the
threshold energy \(E=V\). At threshold one has \(U_{12}=U_{21}=0\) and
\(U_{22}=1\) so \(\xi(V)=0\) but there is no corresponding eigenfunction.
Indeed at this energy the expression for the wavefunction at
\(x>L_1\) contains a division by zero (one may avoid this by normalizing
in a different way but that will destroy unitarity of \(U\)
above threshold which is essential for our approach).\\
Above the critical energy \(E>V\) 
this quantum map is manifestly unitary which describes the flux
conservation at the central vertex. Below the critical value
\(E<V\) the quantum map is not unitary, one may observe however that
\( |U_{11}|=1\) is unimodular as \(k_2=i|k_2|\) in this case. At the
critical value \(E=V\) one has \( |U_{11}|=|U_{22}|=1\).

Using Cauchy's argument principle above threshold \(E>V\) where
\(U(k)\) is unitary
one may then
write the spectral
counting function in the standard way as a trace formula
\begin{subequations}
  \begin{align}
    N_{\mathrm{at}}(E)=
    &
      {\overline{N}}_{\mathrm{at}}(E)+ N_{\mathrm{at}, \mathrm{osc}}(E)
      \label{eq:N_at}
    \\
    {\overline{N}}_{\mathrm{at}}(E)=
    &
      \frac{1}{2\pi} \mathrm{Im} \log \det\left(U(E) \right)+c_{\mathrm{at}}
      \nonumber\\
    =& \frac{L_1 \sqrt{E}}{\pi} +
       \theta\left(E -V\right)\frac{L_2 \sqrt{E-V}}{\pi} -\frac{1}{2}+ c_{\mathrm{at}}
       \label{eq:N_mean_at}
    \\
    N_{\mathrm{at}, \mathrm{osc}}(E)=
    &
      -\frac{1}{\pi}\mathrm{Im} \log
      \det\left(1- U(E+i \epsilon) \right)\nonumber\\
    =
    &
      \sum_{n=1}^\infty
      \frac{1}{n \pi} \mathrm{Im}\ \mathrm{tr}\ U(E+i\epsilon)^n  
      \label{eq:N_osc_at}
  \end{align}
\end{subequations}
where the limit \(\epsilon \to 0\)
from above is implied and the suffix `\( \mathrm{at} \)' refers to
`above threshold'. To facilitate the notation we shall from here on often
omit the reference to the energy dependence and the limit
\(\epsilon \to 0\) of
various quantities in the sequel (writing, for instance, \(U_{11}\) instead of
\( U_{11}(E+i\epsilon) \)).
Note that the traces \(\mathrm{tr}\ U^n\) may be rewritten as sum over
\emph{periodic orbits} \(p\) that visits \(n\) edges
\begin{equation}
  \mathrm{tr}\ U^n = \sum_{p} \frac{n}{r_p} A_p e^{i W_p}\ `
\end{equation}
where the following notation has been used:
a periodic orbit is an equivalence class (with respect to cyclic
permutation)
of a sequence
\( p\equiv \overline{\tau_1\dots \tau_n}\) where each \(\tau_l \in \{1,2\}\) 
corresponds to a section of the orbit which involves the transversal
from the center to the outside vertex and back.
Its length is \( 2L_{\tau_{l}}\). 
The periodic orbit \( p\)
is called \emph{primitive} if the sequence \(p\) is not a
repetition of a shorter
sequence. If \(p\) is not primitive it is the repetition of a
shorter primitive
orbit \(\tilde{p}\)
with \emph{repetition number} \(r_p\). The \emph{scattering amplitude}
of the periodic orbit is given by the product of all scattering amplitudes
collected along the orbit
\begin{equation}
  A_p= (-1)^n \prod_{l=1}^n \sigma_{\tau_{l+1}\tau_{l}} 
\end{equation}
(with the obvious understanding that \(\tau_{n+1}=\tau_1\) as required by periodicity). If \(p\) is not primitive then  \(A_p = A_{\tilde{p}}^{r_p}\).
Finally the phase of the periodic orbit is given by
\begin{equation}
  W_p = 2 n_1 k_1 L_1 +2 n_2 k_2 L_2
\end{equation}
where \(n_1\) and \(n_2=n-n_2\)
are the integer numbers of times \(p\) visits the
corresponding interval
(that is the number of occurrences of the numbers 1 and 2 in the sequence).
Altogether we may then write
\begin{equation}
  N_{\mathrm{at}, \mathrm{osc}}(E)=
  \sum_{\tilde{p}}\sum_{r=1}^\infty \frac{1}{\pi r} \mathrm{Im}\
  A_{\tilde{p}}^r e^{i r W_{\tilde{p}}}
\end{equation}
as a sum over primitive periodic orbits \(\tilde{p}\)
of arbitrary length and their repetitions \(r\).
In the division of the spectral counting function \(N_{\mathrm{at}}(E)=
{\overline{N}}_{\mathrm{at}}(E)+ N_{\mathrm{at}, \mathrm{osc}}(E)\)
the mean part \({\overline{N}}_{\mathrm{at}}(E)\) is a continuous increasing
function of \(E\) while \(N_{\mathrm{at}, \mathrm{osc}}(E)\) is not continuous
(for \( \epsilon=0\)) and oscillating.
Note that all phases \(S_{\tilde{p}}\) are real increasing
functions of \(E\) above threshold. 

The identity \(N(E)= N_{\mathrm{at}}(E)\) is valid above threshold \(E>V\)
for an appropriate choice of the constant \(c_{\mathrm{at}}\) that may be found
if one knows \(N(E)\) at some value \(E > V\). We will show later
that the appropriate choice is \( c_{\mathrm{at}}=0\).
The expression \(N_{\mathrm{at}}(E)\) as a function of \(E\) may be evaluated
also below the critical value \(E<V\)
but
it is not applicable because the derivation assumes that \(U\) is unitary.
Indeed we will show that below threshold \(N_{\mathrm{at}}(E) \neq N(E)\)
and additional
terms appear that vanish above threshold.

So 
let us now derive a trace formula with an alternative approach.
This approach will be valid for all \( E>0\). It is easy to show that the
spectrum is strictly positive, so the whole spectrum is covered.
This approach starts by eliminating the modes in the interval \(x \in
[L_1,L] \) and thus rewriting the quantization condition
\eqref{quantisation_condition}
as
\begin{equation}
  u_\mathrm{red} b_1 =b_1
\end{equation}
where
\begin{equation}
  u_\mathrm{red} = U_{11}+ U_{12}\frac{1}{1-U_{22}}U_{21}
  = \frac{e^{2 i (k_1 L_1 + k_2 L_2)}
    -\frac{k_1-k_2}{k_1+k_2}e^{2i k_1 L_1} }{1-
    \frac{k_1-k_2}{k_1+k_2}e^{2 i k_2 L_2}}
  \ .
\end{equation}
The corresponding reduced secular function is just
\begin{equation}
  \xi_{\mathrm{red}}(E)=1- u_{\mathrm{red}}
  \label{eq:xi_bt}
\end{equation}
and
the energy eigenvalues may be obtained
one-to-one from the condition \(\xi_{\mathrm{red}}(E)=0 \) for the entire
range of \(E\).
It is straight forward computation to prove that
\(u_{\mathrm{red}}\)
is unitary (unimodular) for any real and positive \(E\). Below threshold
\(0<E<V\) one has \(k_2=i|k_2|\) while \(k_1\) is real. In this case
one may write 
\(u_{\mathrm{red}}= - \frac{k_1-i|k_2|}{k_1+i|k_2|} e^{2ik_1L_1}
\frac{1-z^*}{1-z}\) with \(z =\frac{k_1-i|k_2|}{k_1+i|k_2|} e^{-2 |k_2|L_2}\)
in terms of three unimodular factors. For \(E>V\)
both \(k_1\) and \(k_2\) are real and
one may write  \(u_{\mathrm{red}}= e^{2i(k_1L_1+k_2L_2)}
\frac{1-\tilde{z}^*}{1-\tilde{z}}\) with
\(\tilde{z} =\frac{k_1-k_2}{k_1+k_2} e^{2 i k_2 L_2}\) in terms of two
unimodular factors. In either case \( u_{\mathrm{red}}\) is
a product of unimodular factors and thus unimodular itself.
At the threshold \(E=V\) the reduced quantum map is continuous
  and unimodular
  with \(u_{\mathrm{red}}(V)=\frac{1+i k_1 L_2}{1-i k_1 L_2}e^{2ik_1L_1} \).
Therefore one can use Cauchy's argument principle to
express the number counting function as
the trace formula
\begin{subequations}
  \begin{align}
    N_{\mathrm{red}}(E)=
    &
      {\overline{N}}_{\mathrm{red}}(E)+ N_{\mathrm{red}, \mathrm{osc}}(E)
      \label{eq:N_bt}
    \\
    {\overline{N}}_{\mathrm{red}}(E)=
    &
      \frac{1}{2\pi}
      \mathrm{Im} \log u_\mathrm{red} + c_{\mathrm{red}}\nonumber \\
    = &
        \frac{L_1 \sqrt{E}}{\pi} +
        \theta\left(E -V\right)\frac{L_2 \sqrt{E-V}}{\pi}- \frac{1}{2}
        \nonumber\\
    &  -\frac{1}{2\pi } \mathrm{Im} \log (1-U_{22})
      +\frac{1}{2\pi } \mathrm{Im}
      \log \left(1-[U^{-1}]_{22}\right)
      \label{eq:N_mean_bt}
    \\
    N_{\mathrm{red}, \mathrm{osc}}(E)=
    &
      -\frac{1}{\pi}
      \mathrm{Im}\ \log \left(1- u_{\mathrm{red}}\right)
      \nonumber \\
    =
    &
      -\frac{1}{\pi}\mathrm{Im} \log
      \det\left(1- U \right)
      + \frac{1}{\pi}\mathrm{Im} \log
      \left(1- U_{22} \right) \ .
      \label{eq:N_osc_bt}
  \end{align}
\end{subequations}
In the second line of the mean part we have set the constant
\(c_{\mathrm{red}}=-\frac{1}{2}\) by requiring that \(N_{\mathrm{red}}(E)\to 0\)
as \(E\to 0\).
This expression is valid above and below threshold. However,
it will be shown that
the division into oscillating and mean part seems more natural
below threshold.
Let us discuss the expression first above threshold where we will show that
it is consistent with the first approach.
Indeed above threshold the \(U\) is unitary such that the inverse
matrix element is \([U^{-1}]_{22}= U_{22}^*\)
and this implies
\(\frac{1}{2\pi}\mathrm{Im}
\log \left(1-[U^{-1}]_{22}\right)=-\frac{1}{2\pi}\mathrm{Im}
\log \left(1-U_{22}\right)\)
in the mean part 
\({\overline{N}}_{\mathrm{red}}(E)\).
Then
\(
\overline{N}_{\mathrm{red}}(E)=
\overline{N}_{\mathrm{at}}(E) -
\frac{1}{\pi}\mathrm{Im}
\log \left(1-U_{22}\right)
-c_{\mathrm{at}}\)
and
\(N_{\mathrm{red,osc}}(E)=
N_{\mathrm{at,osc}}(E) +
\frac{1}{\pi}\mathrm{Im}
\log \left(1-U_{22}\right)\)
and the total expressions for the number counting function
coincide
\(N_{\mathrm{red}}(E)=N_{\mathrm{at}}(E)\) with the choice
\( c_{\mathrm{at}}=0 \). However the 'mean' and 'oscillating'
parts come out differently as the term
\(-\frac{1}{\pi } \mathrm{Im} \log (1-U_{22})\) has moved from the
oscillating part to the mean part in the decomposition of
\(N_{\mathrm{red}}(E)\). In terms of periodic orbits this corresponds to
the contribution of the primitive orbit \( \tilde{p}= \overline{2} \)
and all its repetitions. For \(E>V\) these contributions are oscillating
functions of \(E\), so the decomposition \(N_{\mathrm{at}}(E)\) of the first
approach seems more natural.
Below threshold \(E<V\) these contributions are no longer oscillating
as the phase \(r W_{\overline{2}}= 2 i r L_2 \sqrt{V-E}\) is purely imaginary
leading to an exponential suppression \(\propto e^{-2 r L_2 \sqrt{V-E}}\)
of these contributions below threshold. So, below threshold it
is indeed natural that these periodic orbit contributions are considered
as part of the mean counting function \({\overline{N}}_{\mathrm{red}}(E)\).
The additional terms in the mean part 
account for the fact that the matrix \(U(E)\) is not unitary
below threshold.
Note that below threshold the inverse matrix element
\begin{equation}
  [U^{-1}]_{22}= \frac{k_1-k_2}{k_1+k_2} e^{-2ik_2 L_2} \equiv
  \frac{k_1-i|k_2|}{k_1+i|k_2|} e^{2 |k_2| L_2}
\end{equation}
becomes exponentially large in modulus.
The logarithms in the mean counting function may then be expanded
with respect to exponentially small terms
\begin{equation}
  -\frac{1}{2\pi } \mathrm{Im} \log (1-U_{22})
  +\frac{1}{2\pi } \mathrm{Im}
  \log \left(1-[U^{-1}]_{22}\right)=
  \frac{1}{2\pi } \mathrm{Im}
  \log \left(-[U^{-1}]_{22}\right)
  +\sum_{r=1}^\infty \frac{1}{2 \pi r}\left(U_{22}^{\ r} - [U^{-1}]_{22}^{-r}
  \right)\ .
\end{equation}
One may interprete the terms in the sum \(\sum_{r=1}^\infty\) 
as the contributions from repetitions of the orbit
\( p=\overline{2}\) where the \(r\)-th repetition
contributes a difference between
the
`standard forward' amplitude \(U_{22}^r\) of the \(r\)-th
repetition using the \(r\)-th power
of the corresponding matrix element of the quantum map
and a `reversed' amplitude \([U^{-1}]_{22}^{-r}\) that uses the
corresponding matrix element of the \textit{inverse} quantum map
(raised to an inverse power).

\begin{figure}[tbh]
  \includegraphics[width=0.75\textwidth]{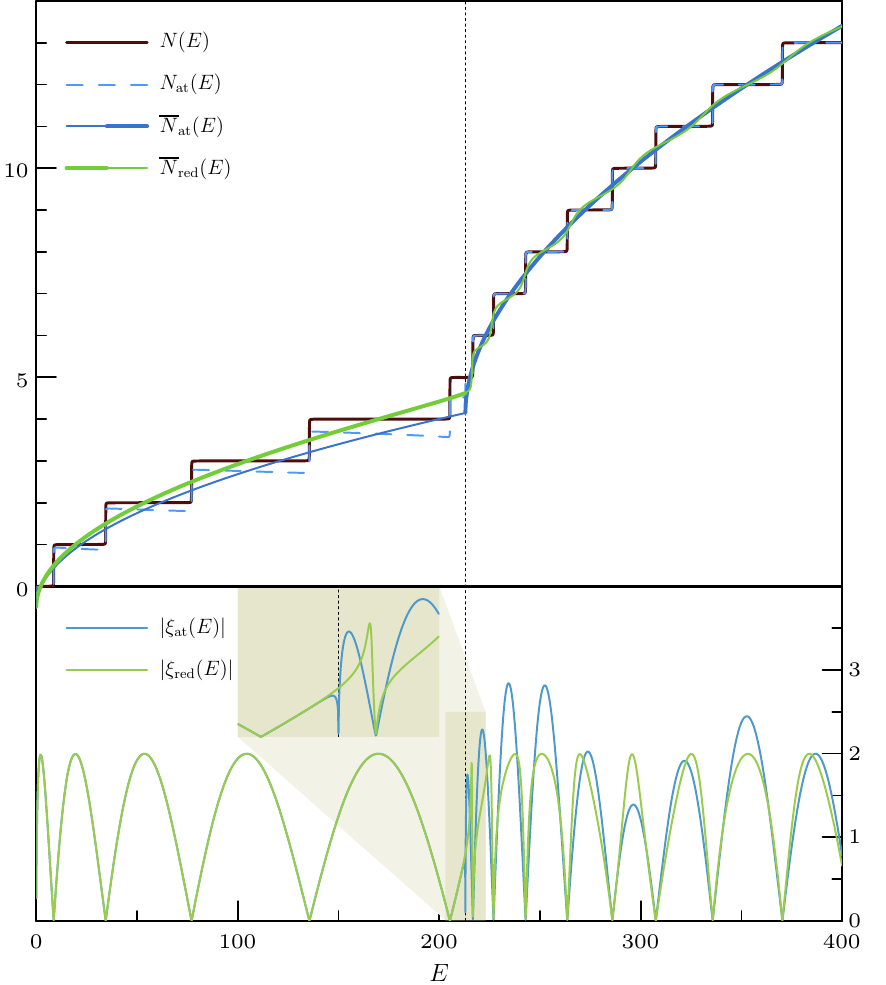}
  \caption{\label{fig1}
    Spectrum for an interval with a potential step in the range
    \(0\le E\le 400\). The lengths were chosen as
    \(L_1=1\) and \(L_2=\sqrt{3}\) and the potential step size
    as
    \(V=213\). The threshold energy \(E=V\) is
    indicated by a dashed vertical line. \\
    Upper panel:
    Exact counting function \(N(E\) (brown staircase) and expression
    \(N_{\mathrm{at}}(E) \)
    (dashed blue line)
    and mean counting functions (full blue and green lines).\\
    Lower panel:
    Secular functions.
  }
\end{figure}

Finally let us discuss the behavior far below threshold by considering
\(E \ll V\) in the asymptotic limit \(V\to \infty\).
In this limit any periodic orbit that visits the interval \( x \in [L_1,L]\)
is suppressed exponentially leaving only contributions from the primitive orbit
\( \tilde{p}=\overline{1} \) and its repetitions
\begin{align}
  {\overline{N}}_{\mathrm{red}}(E)
  \sim
  &
    \frac{L_1 \sqrt{E}}{\pi} -\frac{1}{2} +  \frac{1}{2\pi }
    \mathrm{Im}\log \left(\frac{\sqrt{V-E}+i\sqrt{E}}{\sqrt{V-E}-i\sqrt{E}}\right)
   \\
  N_{\mathrm{red}, \mathrm{osc}}(E)\sim
  &
    -\frac{1}{\pi}
    \mathrm{Im}\ \log \left(1- U_{11}\right)
    \ .
\end{align}
Note that \( U_{11}=-\frac{\sqrt{E}-i \sqrt{V-E}}{\sqrt{E}+i \sqrt{V-E}}
e^{2 i L_1 \sqrt{E}}
\) is unimodular for \(E< V\) and we have only neglected exponentially
small terms while keeping any corrections of order
\(\mathcal{O} \left((E/V)^n \right) \) for arbitrary large \(n\).
Moreover, we have used  \(\frac{1}{2\pi}\mathrm{Im}
\log \left(-[U^{-1}]_{22}\right)=\frac{1}{2\pi}
\mathrm{Im}\log \left(\frac{\sqrt{V-E}+i\sqrt{E}}{\sqrt{V-E}-i\sqrt{E}}\right)\) in the mean counting function. This contribution shifts the counting
function by values
\begin{equation}
  0 \le \frac{1}{2\pi}\mathrm{Im}
  \log \left(
    \frac{\sqrt{V-E}+i\sqrt{E}}{\sqrt{V-E}-i\sqrt{E}}
  \right)\le \frac{1}{2},
\end{equation}
and increases from the lower
bound to the upper bound as \(E\) increases from zero to \(V\).

Let us illustrate this with a numerical example.
In Figure~\ref{fig1}
we plot counting functions in the upper
and the absolute value of the secular function in the lower panel
for a specific (arbitrary) choice of parameters (see figure caption).
The exact counting function \(N(E)\) (fat brown staircase line)
and the
expression \(N_{\mathrm{at}}(E)\) (dashed blue line) given by \eqref{eq:N_at}
only coincide above threshold \(E>V\).
Below threshold  \(N_{\mathrm{at}}(E)\)
shows
steps at the energy eigenvalues but the plateau between steps is not
constant. It has an additional step of half size at the threshold \(E=V\)
This may all be expected from the fact that the secular function
has a spurious zero at threshold and is defined in terms of a unitary
matrix only above threshold.\\
The upper panel also contains plots of the mean counting functions
\(\overline{N}_{\mathrm{red}}(E)\) (green line) and
\(\overline{N}_{\mathrm{at}}(E)\)
(blue line) as given by the trace formulas \eqref{eq:N_mean_bt}
and \eqref{eq:N_mean_at}.
Above threshold both trace formulas coincide for the
full counting function but give different divisions into a `mean' and
`oscillating' part. Comparing the two `mean' parts above threshold
it is apparent that \(\overline{N}_{\mathrm{red}}(E)\) oscillates around
\(\overline{N}_{\mathrm{at}}(E)\). Indeed the difference of the two corresponds to
periodic orbits that remain inside the edge \(x>L_1\) which are
oscillatory functions of the energy for \(E>V\) with an amplitude that
decays with \(E \to \infty\) (when the potential step becomes more and more
transparent). So one may view \(\overline{N}_{\mathrm{at}}(E)\) as the more natural
candidate for the mean part above threshold.
We have therefore plotted \(\overline{N}_{\mathrm{at}}(E)\) with a fatter line in
this region. Below threshold \(E < V\) both \(\overline{N}_{\mathrm{at}}(E)\)
and \(\overline{N}_{\mathrm{red}}(E)\) are smooth increasing functions. However only
\(\overline{N}_{\mathrm{red}}(E)\) is related to an exact trace formula for the
counting function. One can see that the mismatch between the exact
counting function \(N(E)\) and the trace formula \(N_{\mathrm{at}}(E)\)
is due to the fact that the corresponding `mean' part 
\(\overline{N}_{\mathrm{at}}(E)\) is too low by the same amount. The more
natural choice
for the mean part below threshold is clearly \(\overline{N}_{\mathrm{red}}(E)\)
which is therefore drawn with a fat line for \(E<V\).
But this means that the natural choice for the mean part of the counting
function switches from the expression \eqref{eq:N_mean_bt}
for \(\overline{N}_{\mathrm{red}}(E)\) to \eqref{eq:N_mean_at}
for \(\overline{N}_{\mathrm{at}}(E)\) at \(E=V\). The two expressions do not
fit together continuously at threshold however.

The lower panel in Figure~\ref{fig1} shows the absolute value of the
secular functions \(\xi(E)\) and \(\xi_{\mathrm{red}}(E)\) (equations
\eqref{eq:xi_at} and \eqref{eq:xi_bt}).
Away from the threshold their zeros coincide and clearly correspond to the increases of
the counting function.
The ratio of the two satisfies
\begin{equation}
  \frac{\xi(E)}{\xi_{\mathrm{red}}(E)}=1-U_{22}=1-
  \frac{k_1-k_2}{k_1+k_2}e^{i2L_2 k_2}\ .
  \label{eq:xi_ratio}
\end{equation}
At threshold \(U_{22}=1\) and this is the reason for the different behavior
of the two secular functions at this energy (see magnified region of
the plot). Below threshold \(|U_{22}|= e^{-2\sqrt{V-E}L_2}\) and
\eqref{eq:xi_ratio} approaches unity exponentially when \(E\) is decreased
and this can clearly be seen in the plots which lie on top of each other
until one gets close to threshold from below.
Above threshold \(|U_{22}|=\frac{\sqrt{E}-\sqrt{E-V}}{\sqrt{E}+\sqrt{E-V}}\sim
\frac{V}{E}  \) (when \(E\gg V\)). So the ratio also tends to unity when
moving away from threshold but only with a slow \(V/E\) decay and this is
consistent with the plot.

\section{Schr\"odinger operators  on 
  quantum graphs for the piecewise constant potentials and multi-mode models}
\label{sec:graph_operator}

The main aim of the remainder of this manuscript is to derive a generalization
of the trace formulas which were presented in the previous section
to quantum graphs with
piecewise constant potentials or multiple modes.
The details will be given in Sections
\ref{sec:scattering} and \ref{sec:trace}.
Before doing so we shall summarize in this section 
the standard description of PCP quantum graphs 
as self-adjoint metric Schr\"odinger operators on metric graphs 
with appropriate matching conditions
following \cite{schrader, grisha_book, kuchment}.
In most applications of quantum graphs one assumes a zero potential
on the edges. The addition of piecewise constant potentials is straight
forward and therefore readers
that are familiar with quantum graphs may skip
most of this -- or just pick up the notation.
We shall then proceed to describe the relation between the differential
operators in the PCP an MM models.

A quantum graph consists of a metric graph \(\mathcal{G}\) and a
self-adjoint Schr\"odinger operator \(\hat{H}\) in the Hilbert space
\( L^2(\mathcal{G}) \)
of square-integrable functions on \(\mathcal{G}\)
\cite{grisha_book, schrader, kuchment}.
Without loss of generality we assume that the graph is connected.
We allow the graph to have parallel edges and loops but
we will assume here that the metric graph is compact. In that case
the graph has a finite number 
\(N_V\) of vertices and a finite number
\(N_E\) of edges.
Each edge \(e\) has a (finite) length \(L_{e}>0\) and a coordinate
\(x_e\in [0,L_e] \) that describes individual points on the edge
such that \(x_e=0\) and \(x_e=L_e\) are the vertices to which the edge is attached. The explicit choice of direction is arbitrary: \( L_e -x_e\) is as good a
coordinate on \(e\) as \(x_e\).\\
The Schr\"odinger operator \(\hat{H}\)
is defined on a dense subspace
of \( L^2(\mathcal{G})\equiv \oplus_{e \in \mathcal{G}} L^2\left([0,L_e] \right)\)
(to be discussed below). Any \emph{wavefunction}
\(\boldsymbol{\Phi}\),
(that is, any element of this subspace) is a collection of \(N_E\)
complex scalar functions
\(\boldsymbol{\Phi}_e \equiv \phi_e(x_e)\in L^2([0,L_e])\)
and the Schr\"odinger operator acts as
\begin{equation}
  \left(\hat{H} \boldsymbol{\Phi}\right)_e=
  - \frac{d^2\phi_e}{dx_e^2}(x_e)  +V_e \phi_e(x_e)\ .
  \label{eq:edge-laplacian}
\end{equation}
Here \(V_e \in \mathbb{R}\) is a potential that is constant on each edge. 
Piecewise constant potentials can be accommodated by adding vertices
at the positions where the value of the potential changes.
In order for the Schr\"odinger operator to be well-defined
one needs to make sense of the
second derivative in a weak way  \cite{grisha_book}.
For this one needs \(\phi_e(x_e)\) to be a continuous
square-integrable function
which is piecewise differentiable. For the stronger requirement
that \(\hat{H}\) defines a self-adjoint operator one needs additional
matching conditions at the vertices. There is no unique choice
of matching conditions and the most general set of matching conditions
can be described in several equivalent ways. We follow the description
of Schrader and Kostrykin \cite{schrader}.
Let us consider one vertex \(v\) and denote
its degree by
\(d_v\). 
Let  \(\mathbf{S}(v)\) be the \emph{star} of \(v\). By definition this 
is set of edges connected to \(v\) (where any loops
are considered as two independent edges by adding an auxiliary vertex).
Clearly \( \left| \mathbf{S}(v) \right|=d_v\).
For any edge \(e\in \mathbf{S}(v)\) we may assume without loss of generality
that
\(x_e=0\) corresponds to the vertex \(v\) on which we focus.
The matching conditions
are a set of \(d_v\) simultaneous linear relations between the wavefunction
and their
derivatives at \(x_e=0\) for all \(e \in \mathbf{S}(v)\)
\begin{equation}
  \sum_{e'\in \mathbf{S}(v)} \left( A_{ee'}\phi_{e'}(0) + B_{ee'}\frac{d\phi_{e'}}{dx_{e'}}(0)
  \right) =0\ .
  \label{eq:general_matching_condition}
\end{equation}
There are \(d_v\) equations, one for each edge \(e\in \mathbf{S}(v)\).
The coefficients \(A_{ee'}\) and \(B_{ee'}\) form two complex
\(d_v \times d_v\)  matrices \(A\) and \(B\)
for which one requires that \(AB^\dagger= BA^{\dagger}\) is a hermitian matrix
and that the \(d_v \times 2d_v\) matrix \( (A,B) \) has maximal rank
\( d_v\). Note that \( A \mapsto CA\) and \(B \mapsto CB\) for an
invertible matrix \(C\) gives equivalent matching conditions.

If one chooses matrices \(A\) and \(B\) with the above conditions
for each vertex \(v\) in \( \mathcal{G}\) then
the self-adjoint Schr\"odinger operator
\(\hat{H}\) is defined on the dense subspace of \(L^2(\mathcal{G})\)
of piecewise differentiable wavefunctions that satisfy the
corresponding matching conditions at all vertices. 

In sections \ref{sec:scattering} and \ref{sec:trace}
we will consider the eigenproblem
\begin{equation}
  \hat{H} \boldsymbol{\Phi}= E \boldsymbol{\Phi}\ ,
\end{equation}
that is the stationary Schr\"odinger equation on the graph.

In a MM quantum graph the scalar wave function on a given edge is
replaced by a multi-component wavefunction.
The various components describe the transversal modes that may be excited above
a threshold energy.
In the present setting we always assume a finite number of modes. This
is essential to
ensure a discrete energy spectrum. 
Quantum graphs with
infinitely many modes and spectra that contain continuous bands have been
considered
\cite{smilansky_model}.

One arrives at a Schr\"odinger operator on a MM graph
by generalizing on one side
\eqref{eq:edge-laplacian} to the MM setting
by adding a diagonal matrix that includes excitation
energies for each transversal mode. On the other side one generalizes
the matching conditions \eqref{eq:general_matching_condition} by replacing
the matrices \(A_{e,e'}\) and \(B_{e,e'}\) by matrices with elements which carry a double index :
\( A_{(e,m),(e',m')}\) and \(B_{(e,m),(e',m')}\) where  \(e, e' \in \mathcal{S}(v)\) 
identify the interacting edges and the \(m, m'\) identify the interacting
modes. 
We give the details of this description in
Appendix~\ref{app:multi-mode}.

A formally equivalent PCP quantum graph
can be constructed by replacing 
each edge of a MM graph by parallel single-mode
edges of the same length. The details of this
construction can also be found in 
Appendix~\ref{app:multi-mode}.
The main difference between the PCP and MM models
is in different physical choices of
matching conditions at the vertices. 

\section{The scattering approach}
\label{sec:scattering}

The scattering approach to a 
quantum graph with \(N_V\) vertices and
\(N_E\) edges has been a very useful tool
for spectral analysis in the single-mode case without potentials
\cite{KS_traceformula,review}.
There, it leads to an explicit quantization condition in terms
of the zeros of a spectral determinant
\( \xi(E)= \det \left( \mathbb{I} - U(E) \right)=0\) where \(U(E)\)
is a unitary matrix of dimension \(2N_E \times 2N_E\) known as the quantum map.
The quantum map is built up as a product of matrices that describe scattering
at each vertex followed by transport along the edges.
In this section we will generalize the scattering approach to MM
and PCP quantum graphs. The explicit formulation will follow the PCP model
which includes the MM model via the formal equivalence
as was discussed
in the previous section and Appendix~\ref{app:multi-mode}.

In the presence of edge potentials the scattering matrix need not be unitary
as some edges may support evanescent modes. Conservation of probability
currents in this case follows from a well-known symmetry of scattering matrices
in the presence of evanescent modes \cite{weidenmueller} that we will derive
explicitly from the general matching conditions.

\subsection{The vertex scattering matrices and its properties}

Let us consider one vertex \(v\) of degree \(d\).
Without loss of generality we
choose the coordinates \(x_e\) on the adjacent edges
such that \(x_e=0\) is the location of the vertex \(v\)
and we 
enumerate the edges of the graph such that \(e=1,\dots,d\)
correspond to the adjacent edges. Collecting
the wavefunctions
on the adjacent edges in a column vector
\(
  \boldsymbol{\phi} (\boldsymbol{x})=    
  (
    \phi_1(x_1),
    \dots,
    \phi_{d}(x_{d})
  )^T  
\)
where \(\boldsymbol{x}=(x_1,x_2,\dots ,x_{d})\)
is the collection of coordinates 
we may rewrite the matching conditions
\eqref{eq:general_matching_condition} in matrix form as
\begin{equation}
  A \boldsymbol{\phi}(0) + B \boldsymbol{\phi}'(0)=0\ .
  \label{eq:matrix_matching_conditions}
\end{equation}
At a given energy \(E\) the solution of the differential equation
\eqref{eq:edge-laplacian}
can be expressed in terms of plane wave propagating in opposite directions.
Combining these we may write a wavefunction that solves the differential equation on all adjacent edges as
\begin{equation}
  \boldsymbol{\phi}(\boldsymbol{x})=
  \frac{1}{\sqrt{K}}e^{iKX}\boldsymbol{b}_{\mathrm{out}}+
  \frac{1}{\sqrt{K}}e^{-iKX}\boldsymbol{b}_{\mathrm{in}}
  \label{eq:scattering_wf}
\end{equation}
where \(X\) is a diagonal matrix with diagonal \(\boldsymbol{x}\)
and \(K\) is a diagonal matrix \(K_{ee'}= \delta_{e'} K_e\)
with the wavenumbers
\begin{equation}
  K_e=\sqrt{E-V_e}
\end{equation}
for each adjacent edge. Note that \(K_e\ge 0\) for \(E\ge V_e\).
For \(E<V_e\) the wavenumber is imaginary and we choose the convention
\( K_e= i |K_e|\) in this case (positive imaginary part). This choice is
consistent with implicit limits \(\epsilon \to 0^+\) in the energy
\( E \mapsto E +i \epsilon\) that will appear in the next section.
In this case the two solutions are increasing or decreasing
exponential functions.
The factors \(\frac{1}{\sqrt{K}}\) in \eqref{eq:scattering_wf}
normalize the plane wave solutions \( \frac{1}{\sqrt{K_e}}e^{iK_e x_e}\)
to unit probability flux (for \(E>V_e\)). As \(K_e=0\) at \(E=V_e\)
we have to assume that the energy is not equal to any of the potentials
on adjacent edges.
Finally \(\boldsymbol{b}_{\mathrm{in/out}}\) denotes \(d\)-dimensional column
vectors that contain the amplitudes of the incoming and outgoing waves.
Note that the direction of a plane wave implied here is
the direction of the corresponding flow for \(E>V_e\) and the direction
of exponential decay for \(E<V_e\). The matching conditions
\eqref{eq:matrix_matching_conditions} allow us to express the outgoing
amplitudes in terms of the incoming amplitudes as
\begin{equation}
  \boldsymbol{b}_{\mathrm{out}}=
  \sigma(K) \boldsymbol{b}_{\mathrm{in}}
\end{equation}
with the \(d\times d\) \emph{vertex scattering matrix}
\begin{equation}
  \sigma(K)= - K^{1/2} \frac{\mathbb{I}}{A+iBK}
  \left(A-i BK  \right) K^{-1/2}= -\mathbb{I} + 2i K^{1/2}
  \frac{\mathbb{I}}{A+iBK} B K^{1/2} \ .
  \label{eq:vertex_scattering_matrix}
\end{equation}
If all potentials \(V_e\) vanish, one may replace the matrix \(K\)
by the real positive wavenumber \(k=\sqrt{E}\) and the expression reduces
to the well-known formula for the energy-dependent
unitary vertex scattering matrix \cite{schrader} for standard quantum
graphs (with vanishing edge potentials).
If the potentials do not vanish then the vertex scattering matrix is in general
not unitary. One may however express it in terms of the unitary matrix
\begin{equation}
  \mathcal{S}
  \equiv \sigma(\mathbb{I})= -\frac{\mathbb{I}}{A+iB}
  \left(A-i B  \right)= -\mathbb{I} + 2i
  \frac{\mathbb{I}}{A+iB}
  B \ .
\end{equation}
Unitarity of \(\mathcal{S}\) follows straight forwardly from the conditions
that \( (A,B) \) has full rank and that \( A^\dagger B=B^\dagger A\)
is Hermitian. Indeed this is just the scattering matrix without potentials
at energy \(E=1\) (or equivalently if all potentials have the same value and
we take the energy to be one unit above the potential).
The relation between \(  \sigma(K) \)
and \(\mathcal{S} \) may be written as
\begin{equation}
  \sigma(K)= \mathcal{R} +
  \mathcal{T} \frac{\mathbb{I}}{\mathbb{I}+ \mathcal{S} \mathcal{R}}
  \mathcal{S} \mathcal{T}
  \label{eq:vertex_scattering_barrier}
\end{equation}
where
\begin{equation}
  \mathcal{R}=\frac{K-\mathbb{I}}{K+\mathbb{I}}
  \qquad \text{and} \qquad \mathcal{T}=\frac{2 K^{1/2}}{K+\mathbb{I}}\ .
\end{equation}
The relation \eqref{eq:vertex_scattering_barrier} is easily checked
algebraically and
has a straight forward
physical interpretation in terms of potential barriers on each edge
which may be taken from the following sketch:
\begin{equation}
  \includegraphics{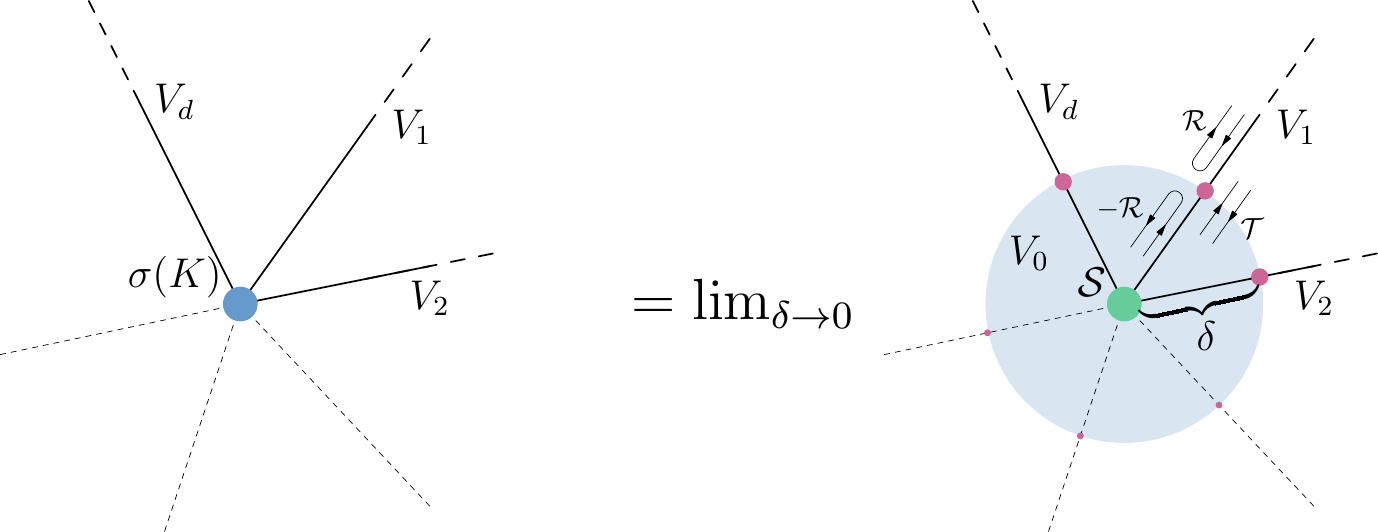}
\end{equation}
For this 
one imagines a small region of size \(\delta>0\) around the
vertex in which the potential has a constant value \(V_0 \equiv E-1\)
and potential 
barriers at the distance \(\delta\). Behind the barrier
the potential is equal to the given edge potential.
The positions of barriers form a set of \(d\) additional vertices.
One then obtains \(\sigma(K)\) as the effective scattering matrix
of the combined barriers and central
vertex with scattering matrix \(\mathcal{S}\) in the limit \(\delta \to 0\)
by observing that \(\mathcal{R}\) is the diagonal matrix of
reflection coefficients
for direct reflection at the barrier without entering the vertex,
\(\mathcal{T}\) is the diagonal matrix of transmission coefficients
across the barrier in either direction, and \( -\mathcal{R}\) gives the
reflection at the barrier from the vertex back into the vertex.
Clearly, \eqref{eq:vertex_scattering_barrier} just describes
the sum of a direct reflection from the barrier plus a term that describes
the transmission through the barrier followed by scattering at the vertex and multiple 
back-reflection into the vertex 
before the final transmission back into the edge.
This is consistent with the scattering matrix
\eqref{eq:example_central_scattering}
at a potential step (considered as a vertex of degree two) in two ways.
On one side
the reflection and transmission coefficients
on the diagonal of \(\mathcal{R}\)
and \(\mathcal{T}\) are obtained from \eqref{eq:example_central_scattering}
at unit energy. On the other side one obtains back
\eqref{eq:example_central_scattering}
at arbitrary energy by using  \eqref{eq:vertex_scattering_barrier} with
\(\mathcal{S} \) as a pure transmission matrix describing continuity of
the wavefunction and its derivative.

For \(E > V_e\) on all adjacent edges \(K\) is a real diagonal matrix
and one finds that \(\sigma(K)\) is unitary.
This can be shown starting from \eqref{eq:vertex_scattering_barrier}
and using the unitary of \(\mathcal{S}\). In general there will be some
edges where \(E < V_e\) and the solutions are evanescent (exponential).
To discuss the structure of the scattering matrix in this case
let us assume that the edges are enumerated such that
\(V_1\le V_2\le \dots V_d\) and consider an energy
\( E \in (V_{e_0}, V_{e_0+1} \) for some edge
\(e_0\in \{1,\dots,d-1\}\) (the cases \(E<V_1\) and \(E> V_d\)
follow straight forwardly from the following discussion).
Then one has oscillatory solutions on the edges on the edges
\(e=1,2,\dots, e_0\) and evanescent solutions on the remaining edges
\(e=e_0+1,\dots,d\). Let us write all matrices as block matrices. For
the vertex scattering matrix one then has
\begin{equation}
  \sigma(K)=
  \begin{pmatrix}
    \sigma(K)_{\mathrm{osc},\mathrm{osc}} & \sigma(K)_{\mathrm{osc},\mathrm{ev}}\\
    \sigma(K)_{\mathrm{ev},\mathrm{osc}} & \sigma(K)_{\mathrm{ev},\mathrm{ev}}
  \end{pmatrix}
\end{equation}
where the index \( \mathrm{osc}\) stands for oscillatory
and \(\mathrm{ev}\) for evanescent.
The diagonal blocks \(\sigma(K)_{\mathrm{osc}, \mathrm{osc}}\)
and  \(\sigma(K)_{\mathrm{ev},\mathrm{ev}}\) are square matrices of
dimension \(e_0 \times e_0\) and \((d-e_0) \times (d-e_0)\).
The other off-diagonal blocks
\(\sigma(K)_{\mathrm{osc}, \mathrm{ev}}\) and \(\sigma(K)_{\mathrm{ev}, \mathrm{osc}}\)
are in general rectangular
of dimension  \(e_0 \times (d-e_0)\) and \((d-e_0)
\times e_0 \). The diagonal matrix \(K\) has real positive elements
on the diagonal in the \(\mathrm{osc}-\mathrm{osc}\) block and positive imaginary entries
in the \(\mathrm{ev}-\mathrm{ev}\) block. Unitarity of \(\mathcal{S}\) and the properties of the
matrix \(K\) lead
to the following symmetry properties for the blocks of
the vertex scattering matrix
\begin{subequations}  \label{eq:vertex_symmetries}
  \begin{align}
    \left(\sigma(K)_{\mathrm{osc},\mathrm{osc}}\right)^\dagger\sigma(K)_{\mathrm{osc},\mathrm{osc}}=
    & \mathbb{I}, \\
    i\left(\sigma(K)_{\mathrm{osc},\mathrm{ev}}\right)^\dagger \sigma(K)_{\mathrm{osc},\mathrm{osc}}=
    &
       \sigma(K)_{\mathrm{ev},\mathrm{osc}}, \\
    i\sigma(K)_{\mathrm{osc},\mathrm{osc}}\left(\sigma(K)_{\mathrm{ev},\mathrm{osc}}\right)^\dagger =
    &
      \sigma(K)_{\mathrm{osc},\mathrm{ev}}, \\
    i \left(\sigma(K)_{\mathrm{osc},\mathrm{ev}}\right)^\dagger\sigma(K)_{\mathrm{osc},\mathrm{ev}}=
    &
      \sigma(K)_{\mathrm{ev},\mathrm{ev}}-
      \left(\sigma(K)_{\mathrm{ev},\mathrm{ev}}^\dagger\right)\ .
  \end{align}
\end{subequations}%
Here the third equation follows directly from the first two equations. The
rest can be found by direct calculation.
These symmetries are well-known in the general context of scattering when
evanescent modes are present \cite{weidenmueller,schanz,rouvinez}.
Each of the four equations can be related to flux conversation \cite{brewer}.
Observing that the outgoing flux on an adjacent edge is given by
\begin{equation}
  I_e=
  \begin{cases}
    |b_{\mathrm{out},e}|^2- |b_{\mathrm{in},e}|^2 & \text{for \(e\le e_0\);}\\
    2\ \mathrm{Im}\  b_{\mathrm{in},e}^* b_{\mathrm{out},e}
    & \text{for \( e > e_0\).}
  \end{cases}
\end{equation}
then the conditions \eqref{eq:vertex_symmetries} ensure
\begin{equation}
  \sum_{e=1}^d I_e =0
\end{equation}
for arbitrary choice of the incoming amplitudes \(\boldsymbol{a}_{\mathrm{in}} \).

\subsection{The quantum map and the quantization condition}

Let us now look at the collection of all vertex scattering matrices
\(\sigma^{(v)}(K)\) for \(v=1,\dots,N_V\) at a given energy \(E\).
We will assume throughout that \(E\) is not equal to any of the
constant potentials on one of the edges.
Each of these matrices
acts on the incoming amplitudes \(\boldsymbol{b}^{(v), \mathrm{in}}\)
of plane waves at the given vertex and
results in the outgoing amplitudes at the same vertex
\(\boldsymbol{b}^{(v), \mathrm{out}}=\sigma^{(v)}(K)
\boldsymbol{b}^{(v), \mathrm{in}}\).
We may collect all
incoming an outgoing amplitudes at all vertices in two
\(2N_E\) dimensional vectors
\(\boldsymbol{b}^{\mathrm{in}}\)
and \(\boldsymbol{b}^{\mathrm{out}}\)
such that each component corresponds to one
directed edge. One may the introduce the \(2N_E\times 2N_E\)
graph scattering matrix
\(S(E)\)
such that
\begin{equation}
  \boldsymbol{b}^{\mathrm{out}}= S(E) \boldsymbol{b}^{\mathrm{in}}\ .
  \label{eq:scattering}
\end{equation}
One can then order the incoming amplitudes in such a way that
the graph scattering matrix is a product of two matrices
\begin{equation}
  S(E)= P \Sigma(K)
\end{equation}
with the block-diagonal matrix
\begin{equation}
  \Sigma(K)=
  \begin{pmatrix}
    \sigma^{(1)}(K) & 0&\dots &0\\
    0& \sigma^{(2)}(K)&\dots&0\\
    \dots &\dots &\dots & \dots\\
    0 & 0 & \dots & \sigma^{(N_V)}(K)
  \end{pmatrix}
\end{equation}
that contains the vertex scattering matrices on the diagonal and a permutation
matrix \(P\). With the convention that we order both amplitude vectors
in the same order with respect to directed edges
(where `in' and `out' give the sense of direction) the permutation matrix
\(P\) just interchanges the two directions on the same edge.
Note that equation
\eqref{eq:vertex_scattering_barrier} remains valid
when replacing \( \sigma(K) \mapsto \Sigma(K)\)
if the matrices \(\mathcal{S}\),
\(K\), \(\mathcal{R}\) and \(\mathcal{T}\) are
extended to  \(2N_E \times 2N_E\) matrices.
Note that the permutation matrix \(P\)
commutes with \(K\), \(\mathcal{R}\) and \(\mathcal{T}\), as these are
diagonal matrices with the same entries for either direction on a given edge.
Reordering the matrix \(\Sigma(K)\) with respect to oscillating and evanescent
modes on the edges for a given energy \(E\) the symmetries
\eqref{eq:vertex_symmetries} also hold for \( \sigma(K) \mapsto \Sigma(K)\).

Next, the local plane wave solutions directly
connect the outgoing amplitude from the start vertex to the incoming amplitude
at the end vertex of a directed edge. This gives the relation
\begin{equation}
  \boldsymbol{b}^{\mathrm{in}}= T(E) \boldsymbol{b}^{\mathrm{out}}\ .
\end{equation}
with the diagonal matrix
\begin{equation}
  T(E)= e^{i K L}
  \label{eq:transport}
\end{equation}
in terms of the two diagonal matrices \(K\) (wavenumbers) and
\(L\) (edge lengths).
The two equations \eqref{eq:scattering}
and \eqref{eq:transport} result in the condition
\begin{equation}
  \boldsymbol{b}^{\mathrm{in}}= U(E) \boldsymbol{b}^{\mathrm{in}}
  \label{eq:quantiszation_condition}
\end{equation}
with the so-called quantum map
\begin{equation}
  U(E)= T(E) S(E)= T(E) P \Sigma(K)=P T(E) \Sigma(K)\ .
  \label{eq:quantum_map}
\end{equation}
In the following the explicit dependence
on the energy \(E\) or the wavenumber matrix \(K\) will often be dropped
for better readability.
The quantization condition may also be written in terms of the secular equation
\begin{equation}
 \xi(E)\equiv \mathrm{det}\left(\mathbb{I} - U(E) \right)=0 
\end{equation}
with the secular function \(\xi(E)\).\\
Let us now fix an energy \(E> \mathrm{min}_{e}(V_{e})\)
and order the directed edges according to increasing potentials.
The corresponding permutation matrix is unitary and thus does not change
the structure of the quantum map \eqref{eq:quantum_map}.
We introduce the oscillatory and evanescent blocks in the same way as in the discussion of
the vertex scattering matrix above:
The directed edges \(e\) where \(E>V_e\) have oscillatory solutions
(superpositions of plane waves) and form the oscillatory subspace where
the remaining edges with \( E < V_e \) form the evanescent subspace
(which may be empty if \(E> \mathrm{max}_e (V_e)\)). Writing all
matrices in block form the quantum map becomes
\begin{equation}
  U\equiv
  \begin{pmatrix}
    U_{\mathrm{osc},\mathrm{osc}} & U_{\mathrm{osc},\mathrm{ev}}\\
    U_{\mathrm{ev},\mathrm{osc}} & U_{\mathrm{ev},\mathrm{ev}}
  \end{pmatrix}\ .
\end{equation}
Then \(U\) inherits from \eqref{eq:vertex_symmetries} the symmetries
\begin{subequations} \label{eq:quantum_map_symmetries}
  \begin{align}
    U_{\mathrm{osc},\mathrm{osc}}^\dagger U_{\mathrm{osc},\mathrm{osc}}=
    & \mathbb{I}\\
    i\left(U_{\mathrm{osc},\mathrm{ev}} \right)^\dagger U_{\mathrm{osc},\mathrm{osc}}=
    & P_{\mathrm{ev},\mathrm{ev}} T_{\mathrm{ev},\mathrm{ev}}^{-1} U_{\mathrm{ev},\mathrm{osc}}\\
    i U_{\mathrm{osc},\mathrm{osc}}\left( U_{\mathrm{ev},\mathrm{osc}}  \right)^\dagger=
    & U_{\mathrm{osc},\mathrm{ev} } P_{\mathrm{ev},\mathrm{ev}}T_{\mathrm{ev},\mathrm{ev}}\\
    i \left(U_{\mathrm{osc},\mathrm{ev}}\right)^\dagger U_{\mathrm{osc},\mathrm{ev}}=
    & P_{\mathrm{ev},\mathrm{ev}} T_{\mathrm{ev},\mathrm{ev}}^{-1} U_{\mathrm{ev},\mathrm{ev}} - U_{\mathrm{ev},\mathrm{ev}}^\dagger P_{\mathrm{ev},\mathrm{ev}} T_{\mathrm{ev},\mathrm{ev}}^{-1}
      \label{eq:quantum_map_symmetries4}
  \end{align}
\end{subequations}
where we have used that the permutation matrix \(P=P^{-1}=P^\dagger\)
is block-diagonal
(as it transposes directions on the same edge) and the \(\mathrm{ev}-\mathrm{ev}\) block
of the transport matrix is real diagonal.\\
If \( E > \mathrm{max}_e ( V_e)\) the quantum map
is unitary. In that case it is straight forward to derive a trace formula
that counts the number of states below a given energy \(E\)
using standard methods. If
\( E < \mathrm{max}_e ( V_e)\) then the quantum map is not unitary and
deriving a trace formula is not as straight forward and it is the topic of the
following section.

\section{The trace formula and its application}
\label{sec:trace}

In the remainder of the manuscript we will focus on
developing a trace formula that counts the number of states below a given
energy \(E\).
For \( E < \mathrm{max}_e ( V_e)\) we will first develop a reduced unitary
description following ideas from \cite{brewer} where analogous considerations
have
been used to deal with evanescent modes in graph-like structures of
mechanical beams.
Once a unitary description is in place we can use standard methods.
We will assume throughout this chapter
that the potentials are ordered \( V_{e} < V_{e+1} \)
and \(E>V_1= \mathrm{min}_e (V_e)\).
Our trace formula will count the number of eigenenergies above
this threshold. This is analogous to the situation in quantum graphs
where the trace formula for the spectral counting
function for a quantum graph with general self-adjoint matching conditions 
\cite{bolte}
only counts the number of states with positive energies while the number of
negative energy states is finite and needs to be determined separately
to obtain the full spectral counting function.

\subsection{The reduced quantum map}

For the given energy \(E\)
we use the corresponding division of the amplitudes
\(\mathbf{b}^{\mathrm{in}}\) in oscillatory and evanescent subspaces.
Equivalently, we can refer to the evanescent and oscillatory subgraph.
Writing the quantization condition in block-forms
\begin{subequations}
  \begin{align}
    U_{\mathrm{osc},\mathrm{osc}} \mathbf{b}^{\mathrm{in}}_{\mathrm{osc}} +
    U_{\mathrm{osc},\mathrm{ev}} \mathbf{b}^{\mathrm{in}}_{\mathrm{ev}}=
    &
      \mathbf{b}^{\mathrm{in}}_{\mathrm{osc}}\\
    U_{\mathrm{ev},\mathrm{osc}} \mathbf{b}^{\mathrm{in}}_{\mathrm{osc}} +
    U_{\mathrm{ev},\mathrm{ev}} \mathbf{b}^{\mathrm{in}}_{\mathrm{ev}}=
    &
      \mathbf{b}^{\mathrm{in}}_{\mathrm{ev}}\ .
  \end{align}
\end{subequations}
Assuming that \(U_{\mathrm{ev},\mathrm{ev}}\) has no unit eigenvalue (see discussion below)
we may rewrite the second equation as
\(
\mathbf{b}^{\mathrm{in}}_{\mathrm{ev}}=\left( \mathbb{I} - U_{\mathrm{ev},\mathrm{ev}}\right)^{-1}
 U_{\mathrm{ev},\mathrm{osc}} \mathbf{b}^{\mathrm{in}}_{\mathrm{osc}}
\)
which allows us to reduce the quantization condition to
a condition on the oscillatory part only
\begin{equation}
  U_{\mathrm{red}}\mathbf{b}^{\mathrm{in}}_{\mathrm{osc}}=\mathbf{b}^{\mathrm{in}}_{\mathrm{osc}}
\end{equation}
with the reduced quantum map
\begin{equation}
  U_{\mathrm{red}}= U_{\mathrm{osc},\mathrm{osc}} +
  U_{\mathrm{osc},\mathrm{ev}}\frac{\mathbb{I}}{\mathbb{I}-U_{\mathrm{ev},\mathrm{ev}} } U_{\mathrm{ev},\mathrm{osc}} \ .
\end{equation}
Physically, flux conservation now requires that the reduced map be unitary
\begin{equation}
  U_{\mathrm{red}}^\dagger U_{\mathrm{red}} = \mathbb{I}.
\end{equation}
Indeed this follows directly from the symmetries 
\eqref{eq:quantum_map_symmetries} between the blocks of the full quantum map.

The determinants of the full quantum map and the reduced quantum map
obey the identities
\begin{equation}
  \frac{\mathrm{det}\ U_{\mathrm{red}}}{ \mathrm{det}\ U}
  = \frac{\det\left(\mathbb{I}- \left(U^{-1}\right)_{\mathrm{ev},\mathrm{ev}}\right)}{
    \det\left(\mathbb{I} - U_{\mathrm{ev},\mathrm{ev}} \right)
  }
  \label{eq:determinant_identity_1}
\end{equation}
and
\begin{equation}
  \mathrm{det}\left(\mathbb{I} - U\right)=
  \mathrm{det}\left(\mathbb{I} - U_{\mathrm{ev},\mathrm{ev}}\right)\
  \mathrm{det}\left(\mathbb{I} - U_{\mathrm{red}}\right)
  \label{eq:determinant_identity_2}
\end{equation}
where \( \left(U^{-1}\right)_{\mathrm{ev},\mathrm{ev}}=\frac{\mathbb{I}}{U_{\mathrm{ev},\mathrm{ev}}
  -U_{\mathrm{ev},\mathrm{osc}}U_{\mathrm{osc},\mathrm{osc}}^{-1} U_{\mathrm{osc},\mathrm{ev}} } \) is the \(\mathrm{ev}-\mathrm{ev}\) block
of the inverse map \(U^{-1}\). These identities
may be derived straight forwardly
purely from the definition of the reduced matrix from the full matrix
(under the assumption that all involved matrices exist).
We will use both identities later to write the trace formula in a
precise yet intuitively appealing way.

Before turning to the trace formula let us comment on the
implicit assumption
that \( \frac{\mathbb{I}}{\mathbb{I}-U_{\mathrm{ev},\mathrm{ev}}}\) exists
in order to
define the reduced map. 
Let us consider in more
detail the situation when this assumption fails
and assume that for some energy \(E\) 
this inverse does not exist.
Identity  \eqref{eq:determinant_identity_2} suggests that
the energy \(E\) is in the spectrum as one of the factors in the secular
equation vanishes. However, the reduced matrix is not defined and one may
question whether the second factor remains finite. So let us
demonstrate more carefully that indeed the energy is in the spectrum.
As \(U_{\mathrm{ev},\mathrm{ev}}\) has (at least one)
unit eigenvalue we may denote the corresponding eigenvector as
\(\boldsymbol{\hat{b}}_{\mathrm{ev}}\).
We claim 
that this eigenvector can be extended to an eigenvector with unit
eigenvalue of
the full quantum map
\begin{equation}
  \begin{pmatrix}
    U_{\mathrm{osc},\mathrm{osc}} & U_{\mathrm{osc},\mathrm{ev}}\\
    U_{\mathrm{ev},\mathrm{osc}} & U_{\mathrm{ev},\mathrm{ev}}
  \end{pmatrix}
  \begin{pmatrix}
    0\\
    \boldsymbol{\hat{b}}_{\mathrm{ev}}
  \end{pmatrix}=
  \begin{pmatrix}
    0\\
    \boldsymbol{\hat{b}}_{\mathrm{ev}}
  \end{pmatrix}\ .
\end{equation}
To prove this one needs to show \(U_{\mathrm{osc},\mathrm{ev}}\boldsymbol{\hat{b}}_{\mathrm{ev}}=0\).
We do this
by
considering the squared norm
\( \left\|U_{\mathrm{osc},\mathrm{ev}}\boldsymbol{\hat{b}}_{\mathrm{ev}}\right\|^2=
\left(\boldsymbol{\hat{b}}_{\mathrm{ev}}\right)^{\dagger} U_{\mathrm{osc},\mathrm{ev}}^\dagger U_{\mathrm{osc},\mathrm{ev}}
\boldsymbol{\hat{b}}_{\mathrm{ev}}=0\)
where the last equality
follows from the symmetry property \eqref{eq:quantum_map_symmetries4}
and using that \(\boldsymbol{\hat{b}}_{\mathrm{ev}}\) is a unit eigenvector of
\(U_{\mathrm{ev},\mathrm{ev}}\).
The extended eigenvector corresponds to a wave function on the graph that
is completely confined to the evanescent subgraph. While this is possible
(e.g. when there are vertices inside the evanescent part with attracting
\(\delta\)-type matching conditions) it requires fine-tuning -- a small
change of edge lengths or matching conditions will deform this eigenstate
to a new one at a shifted energy such that it leaks out into the full graph.
By using the spectral decomposition of \(U_{\mathrm{ev},\mathrm{ev}}\)
near the energy where it is not invertible one can then define
the reduced map \(U_{\mathrm{red}}\) continuously in a neighborhood.
However the identity \eqref{eq:determinant_identity_2} shows
that the reduced secular function
\begin{equation}
  \xi_{\mathrm{red}}(E)=\det\left(\mathbb{I}-U_{\mathrm{red}}(E) \right)
\end{equation}
is generally not zero at an energy \(E\) where \(U_{\mathrm{ev},\mathrm{ev}}\)
has a unit eigenvalue though we have just shown that it is in the spectrum.
A trace formula based on the quantization condition
\(\xi_{\mathrm{red}}(E)=0\) may thus miss some states. For the remainder we will
assume that \(U_{\mathrm{ev},\mathrm{ev}}\) has no unit eigenvalues for any
(relevant) energy. This is indeed generic as a small change
of parameters (lengths, potentials) will immediately lead to some leakage
into the oscillatory part of the graph.
In Section~\ref{sec:star} we construct this situation explicitly for some
example graphs and investigate this numerically.

\subsection{The trace formula}

With a unitary reduced map \(U_{\mathrm{red}}(E)\)
and a quantization condition
\( \mathrm{det}\left( \mathbb{I}- U_{\mathrm{red}}(E) \right)=0\)
Cauchy's argument principle allows us to write the spectral
counting function (or staircase function) as the trace formula
\begin{subequations}
  \begin{align}
    N(E)=
    &
      \overline{N}(E) + N_{\mathrm{osc}}(E)\\
    \overline{N}(E)=
    &
      \frac{1}{2\pi} \mathrm{Im}\ \log \mathrm{det}\left(
      U_{red}(E+i\epsilon)\right) +c
    \nonumber \\
    =&
       \sum_{e=1}^{N_E} \theta(E-V_e) \frac{L_e \sqrt{E-V_e}}{\pi}
      +\frac{1}{2\pi}
      \mathrm{Im}\ \log \mathrm{det}\left(
      S(E+i\epsilon)\right)+c
      \nonumber\\
    &
      +
      \frac{1}{2\pi}
      \mathrm{Im}\ \log \mathrm{det}\left(
      \mathbb{I}- \left(U(E+i\epsilon)^{-1}\right)_{\mathrm{ev},\mathrm{ev}}\right)
      -
      \frac{1}{2\pi}
       \mathrm{Im}\ \log \mathrm{det}\left(
       \mathbb{I}- U_{\mathrm{ev},\mathrm{ev}}(E+i\epsilon)\right)
       \label{eq:weyl}
    \\
    N_{\mathrm{osc}}(E)=
    &
      -\frac{1}{\pi} \mathrm{Im}\ \log \mathrm{det}\left(\mathbb{I}-
      U_{\mathrm{red}}(E+i\epsilon) \right)\nonumber\\
    =&-\frac{1}{\pi} \mathrm{Im}\ \log \mathrm{det}\left(\mathbb{I}-
       U(E+i\epsilon) \right)+
       \frac{1}{\pi} \mathrm{Im}\ \log \mathrm{det}\left(\mathbb{I}-
       U_{\mathrm{ev},\mathrm{ev}}(E+i\epsilon) \right)\nonumber\\
    =&\sum_{n=1}^\infty \frac{1}{n}\left(  \mathrm{tr}\ U^n 
       -\mathrm{tr}\ U_{\mathrm{ev},\mathrm{ev}}^n \right)=
       {\sum_{p}}' \sum_{r=1}^{\infty}
                   \frac{1}{r}  A_p^r e^{i r W_p}
                   \label{eq:periodic_orbit_sum}
  \end{align}
\end{subequations}
which is valid for all energies \(E> V_1=\mathrm{min}_e(V_e)\).
We have used the identities
\eqref{eq:determinant_identity_1} and \eqref{eq:determinant_identity_2}. 
Note that the individual expressions are not continuous
at energies that equal to any potential \(E=V_e\)
as the dimension of the blocks and the reduced map change at these energies.
The constant \(c\) may be evaluated from requiring that
\( \lim_{E\to V_1^+} N(E) \) is equal to the number
of eigenenergies smaller or equal to \(V_1\).
In the oscillatory part we have used
\(  \log \mathrm{det}\ (\mathbb{I}-U )=
\mathrm{tr}\ \log  (\mathbb{I}-U )= - \sum_{n=1}^\infty \frac{1}{n}
\mathrm{tr}\ U^n   \) and wrote the traces as a sum over
\textit{primitive periodic orbits} \(p\) on the graph.
Let us denote a directed edge \(\mathbf{e}\)  as a pair
\( \mathbf{e} \equiv (e,d)\) 
where \(e\) is an edge and \(d=\pm\)
is the direction (for some given
choice of `positive' and `negative' direction).
A periodic orbit of length \(n\) is a  cyclic
sequence \( \overline{\mathbf{e}_1 \mathbf{e}_2
  \dots \mathbf{e}_n} \equiv \overline{\mathbf{e}_2\dots \mathbf{e}_n
  \mathbf{e}_1}\)
of \(n\) directed edges 
such
the vertex at the end  of \(\mathbf{e}_j\) coincides with the vertex of the
start
of \(\mathbf{e}_{j+1}\). Cyclic means \(\mathbf{e}_{n+j}
\equiv \mathbf{e}_{j}\) (the
the start of
\(\mathbf{e}_1\) is the end of \(\mathbf{e}_n\)) and considering an 
equivalence class with respect to the starting edge. The periodic orbit is
primitive if
it is not the repetition of a shorter orbit. The sum
on the right of  \eqref{eq:periodic_orbit_sum} expresses the
oscillatory part of the counting function as a sum of contributions from
primitive periodic orbits
\(p=\overline{\hat{e}_1, \hat{e}_2\dots \hat{e}_{n_p}}\)
of length \(n_p\)
and their repetitions \(r\). To each primitive orbit one associates an
amplitude \( \prod_{j=1}^{n_p} U_{\hat{e}_{j+1} \hat{e}_j}\equiv A_p e^{i W_p}\)
where \(W_p= \sum_{j=1}^{n_p} K_{e_j} L_{e_j} \) and
\(A_p=\prod_{j=1}^{n_p}S_{\hat{e}_{j+1} \hat{e}_j}  \) is the product of scattering
matrix elements.
The prime in the summation indicates that
only primitive orbits that have at least one directed edge in the
oscillatory subgraph contribute, that is the subgraph that consists of all
edges with \(V_e<E\).
These are characterized by \( \mathrm{Re}\
W_p \neq 0\). The contributions from these orbits are oscillatory
because of the factor \( e^{i   \mathrm{Re}\ W_p}\)  which is an
oscillatory function of the energy. The imaginary part of \(W_p\) corresponds
to the evanescent edges that are visited and leads to an exponential
suppression of these orbits due to a factor \( e^{-  \mathrm{Im}\ W_p }\).
At a given energy one may distinguish three types of orbits \(p\):
either all edges of \(p\) are in the evanescent part
(for these orbits \(\mathrm{Re}\ W_p=0\)),
or all edges of \(p\) are in the oscillatory part
(in this case \(\mathrm{Im}\ W_p=0\)) or \(p\) visits both the
evanescent and the oscillatory subgraphs. Only the latter two types
have are contained in the oscillatory part of the counting function,
and far below the next critical energy the
orbits that are purely oscillatory orbits are dominant.
We will show below that one part of the mean counting function
\eqref{eq:weyl} contains contributions from
purely evanescent periodic orbits.

One may wonder why the constant \(c\) has the same value
when the individual parts of the expression are not continuous at \(E= V_e\).
Should one not choose different constants in each interval such that the
counting function remains continuous at these energies
(or jumps by an integer if they happen to be in the spectrum).
The reason for this lies in the fact that there is an element of choice
in the formula that we have given.
E.g. the reduced map as we have defined it has dimension two for
energies \(V_1< E <V_2\) and then changes to dimension four in the interval
\(V_2 < V_3\) and so forth. Alternatively one may stick to a reduced map
of dimension two for all energies \(E>V_1\) without the restriction
\(E<V_2\). When crossing \(E=V_2\) the two-dimensional reduced map remains
unitary and the trace formula remains valid. Just the designation of the
blocks as oscillatory and evanescent becomes blurred as the \(\mathrm{ev}-\mathrm{ev}\) block now
acts on a subgraph that has one oscillatory edge. The unitarity of this matrix
across such a crossing can be shown explicitly and follows directly from the
fact
that one may reduce the matrix in steps and reducing an already
unitary further will always lead to a smaller unitary matrix.
Analogously the formula with a reduced matrix of given size
\(2e \times 2e\) is valid for all energies \( E> V_e\).
Eventually for \(E>V_{N_E}\) one may use any of the \(N_E-1\)
reduced matrices, or the full matrix \(U\).
This does not imply that the individual terms \(\overline{N}(E)\)
and \(N_{\mathrm{osc}}(E)\) are the same for all these choices -- only their
sum is not affected by this choice. This can be seen directly if we fix
the dimension of the reduced matrix but consider the expression at energy
\(E>V_{N_E}\). At this energy the full matrix has become unitary
\(U^{-1}= U^\dagger\) such that \((U^{-1})_{\mathrm{ev},\mathrm{ev}}=(U^\dagger)_{\mathrm{ev},\mathrm{ev}}=
U_{\mathrm{ev},\mathrm{ev}}^\dagger\). In that case the third term in the
expression for \(\overline{N}(E)\)
obeys
\(\mathrm{det}\left(
      \mathbb{I}- \left(U^{-1}\right)_{\mathrm{ev},\mathrm{ev}}\right)
= \mathrm{det}\left(
  \mathbb{I}- U^{\dagger}_{\mathrm{ev},\mathrm{ev}}\right)=
\mathrm{det}\left(
  \mathbb{I}- U_{\mathrm{ev},\mathrm{ev}}\right)^*\) which implies that the
third and forth term can be combined
to \(\frac{1}{\pi} \mathrm{Im}\ \log \mathrm{det}\left(\mathbb{I}-
  U_{\mathrm{ev},\mathrm{ev}}\right)\)
which appears with the opposite sign in \( N_{\mathrm{osc}}(E)\).
When looking at the complete counting function these terms then cancel and what
remains is just the expression one would have obtained directly from
full matrix. This identity however works only if the constant term \(c\)
is also the same in both the expressions.

The main reason why it seems more natural to let the dimension of the 
reduced map increase
by two at each energy \(E=V_e\) rather than just use the trace formula with a reduced
map of dimension two throughout all energies \(E>V_1\)
(or another fixed even dimension above a corresponding threshold energy) is that in the latter case
the division between oscillatory and evanescent subgraph does not correspond
to the periodic orbits that contribute in the oscillatory part of the counting functions.
So let us assume again that the energy is in one interval
\(V_e < E< V_{e+1}\). Above we have already shown that the
oscillatory part of the counting function can be written as a sum over periodic orbits that
are either purely oscillatory or visit both  the oscillatory and the evanescent subgraphs.
These are the orbits whose contributions show
strongly oscillatory behavior as functions of energy
because \( \mathrm{Re}\ W_p \) is an increasing function of the energy. Let us now come back, as promised above,
to the fate of the purely evanescent orbits. In the expression for the oscillatory part of
 	\ref{eq:periodic_orbit_sum} they are explicitly subtracted via the term
 \begin{equation}
 \frac{1}{\pi}
       \mathrm{Im}\ \log \mathrm{det}\left(
       \mathbb{I}- U_{\mathrm{ev},\mathrm{ev}}\right)= - \sum_{n=1}^\infty \frac{1}{n \pi}   
       \mathrm{tr}\ U_{\mathrm{ev},\mathrm{ev}}^n
 \end{equation}
One half of this term reappears with the opposite sign in the mean part. The missing half
appears	 in a different form as \(\frac{1}{2\pi}\
      \mathrm{Im}\ \log \mathrm{det}\left(
      \mathbb{I}- \left(U^{-1}\right)_{\mathrm{ev},\mathrm{ev}}\right)\).
The latter cannot be expanded directly into traces of powers of 
\(\left(U^{-1}\right)_{\mathrm{ev},\mathrm{ev}}\) because this matrix contains exponentially large entries
\( \propto T^{-1}_{\mathrm{ev},\mathrm{ev}}\). 
Factoring out the matrix one may expand the two logarithmic determinants in the mean part as
\begin{equation}
\begin{split}
\frac{1}{2\pi}	
      \mathrm{Im}\ \log \mathrm{det}\left(
      \mathbb{I}- \left(U^{-1}\right)_{\mathrm{ev},\mathrm{ev}}\right)
      -
      \frac{1}{2\pi}
       \mathrm{Im}\ \log \mathrm{det}\left(
       \mathbb{I}- U_{\mathrm{ev},\mathrm{ev}}\right)
       = \\
       \frac{1}{2\pi} \log\det\left( - \left(U^{-1}\right)_{\mathrm{ev},\mathrm{ev}}\right)
       +\sum_{n=1}^\infty \frac{1}{2\pi n}\left( \mathrm{tr}\ U_{\mathrm{ev},\mathrm{ev}}^n -
       \mathrm{tr}\ \left(U^{-1}\right)_{\mathrm{ev},\mathrm{ev}}^{-n}
       \right)
       \end{split}
\end{equation}
where the terms sum over traces may be expanded further into contributions over purely evanescent
periodic orbits such that each orbit has two amplitudes one standard
contribution where amplitudes are
products of matrix elements of the quantum map and a second
`reversed' contribution from the negative powers
of the \(\mathrm{ev}-\mathrm{ev}\)-block of the inverse quantum map.
Both contributions are exponentially suppressed when one is well below the next energy threshold \( E\ll V_{e+1}\).
The first term \(\frac{1}{2\pi}
       \mathrm{Im}\ \log \mathrm{det}\left(
       \mathbb{I}- U_{\mathrm{ev},\mathrm{ev}}\right)\) contributes a term of order unity in the 
       whole interval \( V_e < E < V_{e+1}\).

\subsection{Example: the star graph with Robin-conditions}
\label{sec:star}

\begin{figure}[h]
  \includegraphics[width=0.85\textwidth]{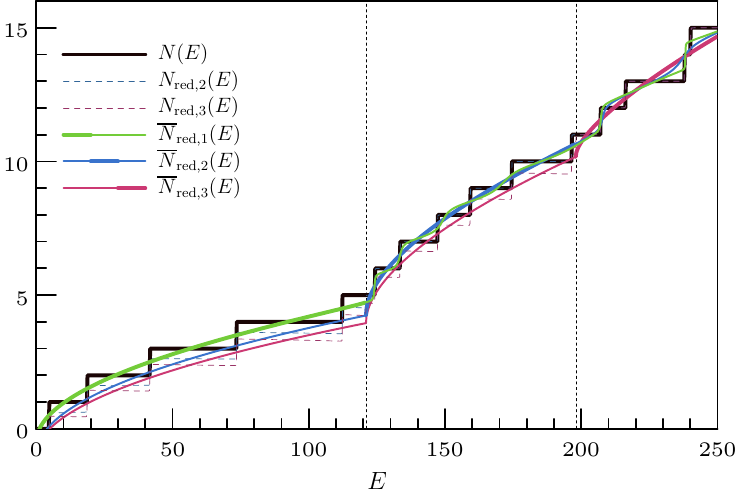}
  \caption{\label{fig2} Counting function for a 3-star with
    \(L_1=\sqrt{2}\), \(L_2=\sqrt{3}\), \(L_3=1\), \(V_1=0\),
    \(V_2=121\), \(V_3=198\) and Dirichlet conditions at all
    vertices of degree one. The exact counting function \(N(E)\)
    (fat brown line) is obtained from finding zeros of the
    secular function. \(N(E)_{\mathrm{red}, n}\) (dashed lines)
    gives the trace formula based on a reduced quantum map of
    dimension \(2n\) (for \(n=3\) this is the full map, \(n=1\)
    coincides with the exact stair case), and
    \(\overline{N}(E)_{\mathrm{red}, n}\) (full lines)
    gives the `mean' parts of these
    trace formulas. The latter are plotted fat in the intervals
    \(V_{n-1}<E<V_n \) 
    where
    the corresponding full trace formula is valid and the
    split into mean part and oscillating part is most natural.
  }
\end{figure}

\begin{figure}[h]
  \includegraphics[width=0.85\textwidth]{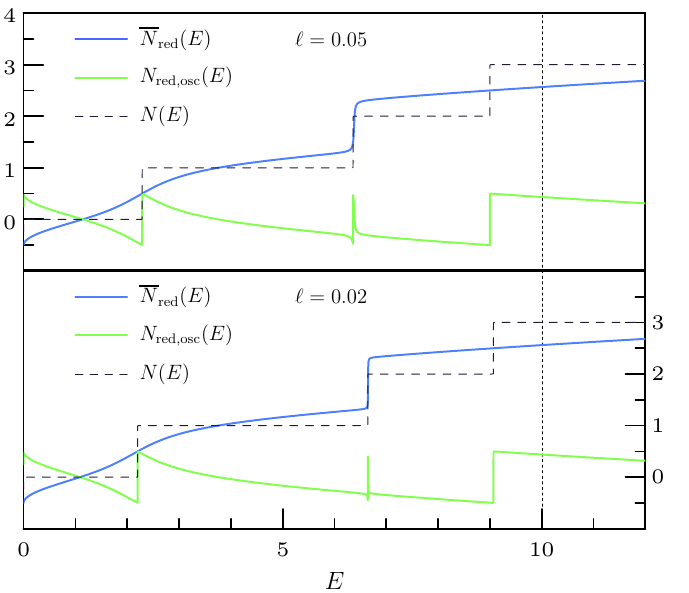}
  \caption{\label{fig3} Counting function
    (mean part \(\overline{N}(E)\),
    oscillating part \(N_{\mathrm{osc}}(E)\), and their
    sum \(N(E)\)
    for a 3-star graph
    with lengths \(L_1=1\), \(L_2=0.5\), \(L_3=L_2+\ell\), \(V_1=0\),
    \(V_2=V_3=10\) with Kirchhoff matching conditions at the center,
    Dirichlet conditions at the end of edge \(e=1\) and
    (attractive) Robin conditions
    with coupling parameters \(\lambda_2=\lambda_3=-2.5\).
    All expressions use the reduced approach appropriate for
    \(E<V_{2/3}\).
    The length mismatch is \(\ell=0.05\) in the upper panel and
    \(\ell=0.02\) in the lower panel.
  } 
\end{figure}

A star graph has one central vertex and \(N_E=N_V-1\) edges that connect the
central vertex to dangling vertices of degree one.
The above description results
in a quantum map \(U(E)\) of dimension \(2N_E\times 2 N_E\) where each
dimension of the map correspond to a directed edge.
The simple topology of star graphs
implies that a plane wave that moves out from the center is reflected
back \textit{on the same edge} in the opposite direction.
As a consequence one often uses an equivalent
description using a smaller quantum map \(\tilde{U}(E)\) that has dimension
\(N_E \times N_E \) where each dimension
corresponds to an \textit{undirected}
edge.
In this case one may write
\begin{equation}
  T = \begin{pmatrix}
    \tilde{T} & 0 \\
    0 & \tilde{T}
  \end{pmatrix}
\end{equation}
and
\begin{equation}
  S= \begin{pmatrix}
    0 & \tilde{\sigma}\\
    \sigma^{(N_E+1)} & 0
  \end{pmatrix}
\end{equation}
where
\(\sigma^{(N_E+1)}\) is the vertex scattering matrix at the center,
\(\tilde{\sigma}=\mathrm{diag}(\sigma^{(1)},\dots,\sigma^{(N_E)} )\) contains
the
scalar scattering coefficients at the dangling vertices, and
\(\tilde{T}=
\mathrm{diag}(e^{i K_1 L_1},\dots , e^{i K_{N_E} L_{N_E}})\) is the diagonal
\(N_E \times N_E\) matrix that contains phase factors for
traveling along one end to the other on each edge.
As the quantum map \(U\) has a block form that vanishes on the diagonal
one then finds that the secular function may be written as
\begin{equation}
  \det\left( \mathbb{I}- U\right) = \det\left( \mathbb{I}- \tilde{U}\right) 
\end{equation}
with
\begin{equation}
  \tilde{U}=\tilde{T} \tilde{\sigma}\tilde{T} \sigma^{(N+1)}=
  \tilde{\sigma}\tilde{T}^2 \sigma^{(N+1)}\ . 
\end{equation}
The matrix \(\tilde{U}\) is a quantum map defined on edges rather than
directed edges and it describes the scattering on incoming plane waves
at the center using
\(\sigma^{(N+1)} \) followed by propagation along the edges from the center
out using \(\tilde{T}\), the reflection at the dangling vertices using
\(\tilde{\sigma}\) and propagation back to the center using
\(\tilde{T}\). Note that \(U\) is unitary if and only if \(\tilde{U}\)
is unitary.

Our introductory example in Section~\ref{sec:interval} can be considered
as the simplest incarnation of a star with \(N_E=2\) edges
corresponding to the two
sub-intervals and Dirichlet conditions.
There we
have used the smaller \(2 \times 2\) description which
is more compact but wrote \(U\) rather than \(\tilde{U}\).
For general graph topologies the description has to be based on directed
edges and that is what we have sticked to in the rest of the description.
For star graphs
is
is straight forward to translate all results obtained using \(U\)
in terms of \(\tilde{U}\).

To illustrate the theory and, especially, how the trace formula
can be applied in practice let us consider a star graph with \(N_E\)
edges and 
assume Kirchhoff
matching conditions at the center (continuity of the wavefunction
through the vertex and a vanishing sum over all edges of the
outward derivative of the wavefunction at the center). 
On the dangling
vertices of degree one we will put
\(\delta\)-type conditions with coupling parameter
\(\{\lambda_e\}_{e=1}^{N_E}\). The latter are also known as Robin-conditions
and are defined by \(\phi_e'(L_e) = \lambda_e\phi_e(L_e) \).
With \(\lambda_e \to \infty\) or \(\lambda_e=0\)
this includes Dirichlet or Neumann conditions
as special cases. 
With our introductory example in Section~\ref{sec:interval}
we have already considered star with \(N_E=2\) edges corresponding to the two
sub-intervals and Dirichlet conditions.
For a three-star, \(N_E=3\) with Dirichlet conditions at all degree one
vertices and some arbitrary choice
of lengths and potentials
one finds similar behavior as for the introductory two-star example, see
Figure~\ref{fig2}.
Apart from having two threshold energies instead of one, we may refer to
the discussion in Section~\ref{sec:interval} of Figure~\ref{fig1}.

One may wonder how the trace formula works when there are eigenstates that
vanish exactly on a subgraph with low edge potentials.
How can the reduced scattering approach `see' these states? Or are they
missed out?
For special choices of the parameters and using Robin conditions with negative
(attracting) coupling parameters one may consider these questions for a
star graph with \(N_E=3\) edges.
To construct such a case, let us
choose \(V_1=0\), \(V_2=V_3>0\),
\(L_2=L_3 \), \(\lambda_1 \to \infty\) and \(\lambda_2=\lambda_3\)
such that the edges \(e=2\) and \(e=3\) are identical.
In that case the eigenstates will either be odd or even
under exchange of the two edges and all odd eigenstates will vanish
on the edge \(e=1\). Choosing Dirichlet conditions everywhere
(that is sending \(\lambda_2=\lambda_3\) to infinity) the odd eigenstates can
be constructed explicitly as \(\phi_1(x_1)=0\),
\(\phi_2(x_2)=A\sin(n\pi x_2/L_2)\) and \(\phi_3(x_3)=-A\sin(n\pi x_3/L_3)\)
for some amplitude \(A \neq 0\) and positive integer \(n\). The corresponding
eigenenergies are \( E= \frac{n^2 \pi^2}{L_2^2}+V_2 >V_2\). For finite
(positive or negative) values of \(\lambda_2=\lambda_3\) these energies
decrease as the coupling parameters are lowered. For Neumann conditions
\(\lambda_2=\lambda_3=0\) they have decreased to
\( E= \frac{(n-1/2)^2 \pi^2}{L_2^2}+V_2 >V_2\). For negative coupling parameter
one may drive the lowest of these energies below the threshold \(V_2\).
As long as the graph is completely symmetric the wavefunction does not leak
into the edge \(e=1\). Let us consider how this situation may be approached
numerically by introducing a small mismatch of the lengths \(L_3=L_2+\ell\).
This is the regime shown in Figure~\ref{fig3} where we plot the counting
function below the lowest threshold. With the given choice of parameters
there are three eigenvalues below threshold. By construction the
wavefunction of the central one becomes completely localized on edges
\(e_2\) and \(e_3\) as \(\ell \to 0\). Plotting the mean and oscillating parts
as defined by the reduced quantum map of dimension \(2 \times 2\) one
can see that the mean part remains smooth at the lower and upper eigenenergy as
\(\ell \to 0\) while it develops a discontinuous step at the central energy.
At the same time the step in the oscillating part narrows to a tiny resonance
at this position (while the steps remain clear for the other two
eigenenergies). If one sets \(\ell=0\) from the start then the trace formula
misses the central eigenenergy: the expressions for smooth and oscillating
part are just continuous here. The limit \(\ell \to 0 \) however creates a step
-- this is possible due to the multi-valuedness of the complex logarithm.
While we have excluded this case by assumption in our derivation,
this numerical
analysis gives an indication that one may define a trace formula with the
reduced quantum map that does not miss out any states that localize in the
evanescent part by using continuity with respect to some parameters
(lengths, potentials, matching conditions).

\section{Outlook}
\label{sec:conclusion}

We have expanded the spectral theory of quantum graphs by
constructing a scattering approach for quantum graphs with piecewise constant
potentials and a multi-mode wave function with a finite number of modes on
each edge. In this finite case it is formally
sufficient to just consider single-mode
quantum graphs with edge-wise constant potentials as one can always map the
multi-mode graph to an equivalent
larger graph with parallel edges, single-mode
wavefunctions and inferred matching conditions.
The presence of evanescent modes involves non-unitary scattering matrices as a
direct consequence. This is a challenge for the construction of
a trace formula for the spectral counting function and we have overcome
this challenge by introducing a reduced unitary approach.\\
The scattering approach for quantum graphs may be used in other settings
straight forwardly, e.g. for scattering
from a finite (compact) graph with a finite number of leads attached.
Many of our constructions are valid beyond quantum graph theory as they
build on the generic symmetries of scattering matrices in the presence of
evanescent modes -- e.g. in the semiclassical scattering approach to
quantum billiards where evanescent modes are always present and there is an
infinite series of energy thresholds where single evanescent modes become
oscillatory.\\
Finally, in a way the trace formula we have presented is not quite complete.
We have assumed that the energy is always larger than the lowest edge
potential. But how do we count the number of states below the lowest potential.
The reduced scattering matrix has zero dimension, so the approach does
not seem to make immediate sense. We leave this open for further
investigation.

\acknowledgments
We would like to thank Stephen Creagh and Gregor Tanner for
helpful discussions. SG thanks for support by COST action CA18232. US
thanks the Department of Mathematical Sciences in the University of Bath
for the hospitality and support and for the nomination as a David Parkin
professor.

\appendix

\section{Multi-mode quantum graphs and their formal equivalence to a
  single-mode quantum graph}
\label{app:multi-mode}

Let us consider the MM setting on a connected metric graph
with \(N_V\) vertices and \(N_E\) edges.
In this setting the scalar wavefunction
\(\phi_e(x_e)\) on the edge \(e\)
is replaced by a multi-component wavefunction
\begin{equation}
  \boldsymbol{\phi_e}(x_e)=
  \begin{pmatrix}
    \phi_{e,1}(x_e)\\
    \dots\\
    \phi_{e,\mu_e}(x_e)
  \end{pmatrix}
\end{equation}
with \(\mu_e\) components and a Schr\"odinger operator \(\hat{H}\)
acts on a given edge as
\begin{equation}
  \left(\hat{H} \boldsymbol{\Phi}\right)_e=
  - \frac{d^2\boldsymbol{\phi_e}}{dx_e^2}(x_e)
  +\boldsymbol{V}_e \boldsymbol{\phi_e}(x_e)
\end{equation}
where diagonal (constant) matrix
\(\boldsymbol{V}_e= \mathrm{diag}\left(
  V_{e,1},\dots,V_{e,\mu_e} \right) \) 
replaces the scalar potential.
We will always assume that the number of modes \(\mu_e\)
is finite on each edge but we do allow \(\mu_e< \infty\) to vary
from one edge to another.
By straight forward extension of \cite{schrader} matching conditions
that render the corresponding Schr\"odinger operator \(\hat{H}\)
self-adjoint follow the same
pattern as in the single-mode case. 
At a given vertex \(v\)of degree
\(d_v\) one may write the matching conditions as
linear relations between the adjacent multi-component wavefunctions
and their derivatives
\begin{equation}
  \sum_{e'} \left( A_{ee'}\boldsymbol{\phi}_{e'}(0) +
    B_{ee'}\frac{d\boldsymbol{\phi}_{e'}}{dx_{e'}}(0)
  \right) =0   
\end{equation}
where the sum is over all edges \(e'\) adjacent to \(v\) and
there are \(d_v\) -- one for each adjacent edge \(e\).
The coefficients \(A_{ee'}\) and \(B_{e e'}\)
are now matrices of size \(\mu_e \times \mu_{e'}\).
Let \(\tilde{d}_v= \sum_{e=1}^{d_v} \mu_e\) (the number of all modes on adjacent edges) then we can combine the coefficient matrices to a large matrix of size
\(\tilde{d}_v\times \tilde{d}_v\) and the linear relations define
a self-adjoint problem if and only if 
\(AB^\dagger= BA^{\dagger}\) is a hermitian matrix
and that the \(\tilde{d}_v \times 2\tilde{d}_v\)
matrix \( (A,B) \) has maximal rank
\( \tilde{d}_v\). If \(\mu_e=1\) on all edges our description of
a multi-mode graph reduces to a single-mode quantum graph with
constant potentials as a special case.
However we may also view a multi-mode graph with \(N_V \)
vertices
with degrees \(\{d_v\}_{v=1}^{N_V}\) and \(N_E= \frac{1}{2} \sum_{v=1}^{N_V} d_v \) edges with
\( \{\mu_e\}_{e=1}^{N_E} \) modes as a single-mode PCP quantum graph
with the same number of vertices \(N_V\)
and \( \tilde{N}_E = \sum_{e=1}^{N_E} \mu_e =\frac{1}{2} \sum_{v=1}^{N_V} \tilde{d}_v\)
single-mode edges by replacing each edge
\(e\) in the original multi-mode graph
by \(\mu_e\) parallel edges of the same length \( L_e\)
with single-mode wave functions \(\phi_{e,m}(x_{e}) \mapsto
\phi_{e,m}(x_{e.m})\) (where \(x_{e,m}\) with \(1\le m \le \mu_e\)
is the coordinate on the \(m\)-th parallel edge).
The
excitation energies \(V_{e,m}\) (\(1\le m \le \mu_e\))
then become constant potentials on the \(m\)-th parallel edge and the
description of self-adjoint matching conditions carries over in a natural way.
In the rest of the paper we will use the formal language of single-mode
PCP
quantum graphs and think of MM quantum
graphs as a special case with parallel edges of the same length.
While this equivalence between the MM and PCP picture
on an enlarged graph is formal it is straight forward
to adapt in a theoretical setting as well as
practically in an experiment. In the former one may prescribe matching
conditions and excitation energies as required and in the latter
the relevant parameters can be measured (or chosen consistently with
available measurements).


\begin{thebibliography}{99}
\bibitem{roth} J.-P. Roth, \textit{Spectre du laplacien sur un graphe},
  C.R.Acad.Sci.~Paris~S{\'e}r. I Math. \textbf{296}, 793–795 (1983).
\bibitem{KS_traceformula}
  T.~Kottos, U.~Smilansky, \textit{Quantum Chaos on Graphs}, Phys.~Rev.~Lett.
  \textbf{79}, 4794 (1997).
\bibitem{review} S.~Gnutzmann, U.~Smilansky,
  \textit{Quantum graphs: Applications to quantum chaos and universal spectral statistics}, Advances in Physics \textbf{55},527 (2006).
\bibitem{grisha_book}
  G.~Berkolaiko, P.~Kuchment, \textit{Introduction to Quantum Graphs}
  (AMS, Providence, 2013).
\bibitem{exp_hul}
  O.~Hul, S.~Bauch, P.~Pako{\'n}ski, N.~Savytskyy, K.~{\.Z}yczkowski,
  L.~Sirko, \textit{Experimental simulation of quantum graphs by microwave networks}, Phys.~Rev.~ E \textbf{69}, 056205 (2004).
\bibitem{exp_allgaier}
  M.~Allgaier, S.~Gehler, S.~Barkhofen, H.-J. St{\"o}ckmann, U.~Kuhl,
  \textit{Spectral
    properties of microwave graphs with local absorption}, Phys.~Rev. E
  \textbf{89}, 022925 (2014).
\bibitem{exp_rehemanjiang}  
  A.~Rehemanjiang, M.~Allgaier, C.H.~Joyner, S.~M{\"u}ller, M.~Sieber,
  U.~Kuhl, H.-J. St{\"o}ckmann,
  \textit{Microwave realization of the gaussian symplectic ensemble},
  Phys.~Rev.~Lett. \textbf{117}, 064101 (2016).
\bibitem{exp_dietz}
  B.~Dietz, V.~Yunko, M.~Bia{\l}ous, S.~Bauch, M.~{\L}awniczak, L.~Sirko,
  \textit{Nonuniversality in the spectral properties of
    time-reversal-invariant microwave networks and quantum graphs},
  Phys.~Rev.~E \textbf{95}, 052202 (2017). 
\bibitem{exp_fu}
  Z.~Fu, T.~Koch, T.M.~Antonsen, E.~Ott, S.M.~Anlage,
  \textit{Experimental Study of Quantum Graphs with
    Simple
    Microwave Networks: Non-Universal Features},
  Acta Physica Polonica A, \textbf{132}, 1655 (2017).
\bibitem{johnson} A.~Johnson, M.~Blaha, A.E.~Ulanov, A.~Rauschenbeutel, 
P.~Schneeweiss, J.~Volz,
\textit{Observation of Collective Superstrong Coupling of Cold Atoms to a 30-m Long Optical Resonator} Phys. Rev. Lett.  \textbf{ 123}, 243602 (2019).
\bibitem{exp_lawniczak}
  M.~{\L}awniczak, P.~Kurasov, S.~Bauch, M.~Bia{\l}ous, V.~Yunko,
  L.~Sirko,
  \textit{Hearing Euler characteristic of graphs},
  Phys.~Rev.~E \textbf{101}, 052320 (2020).
\bibitem{exner_book} P.~Exner, H.~Kovarik, \textit{Quantum Waveguides}, (Springer, Chams, 2015).
\bibitem{olaf_book} O.~ Post,
  \textit{Spectral Analysis on Graph-like Spaces} (Springer, Berlin, 2012).
\bibitem{weidenmueller}
  H.A.~Weidenm\"uller,  \textit{Studies of Many-Channel Scattering},
  Annals of Physics \textbf{28}, 60-115 (1964).
\bibitem{schanz}
  H.~Schanz, U.~Smilansky, \textit{Quantization of Sinai's billiard -- a scattering approach}, Chaos, Solitons \& Fractals \textbf{5}, 1289-3009 (1995)
\bibitem{rouvinez} C.~Rouvinez, U.~Smilansky, \textit{A scattering approach to the quantization of Hamiltonians in two dimensions -- application to the wedge
    billiard}, J.~Phys. A \textbf{98}, 77-104 (1995).
\bibitem{brewer} C.~Brewer, S.C.~Creagh, G.~Tanner,
  \textit{Elastodynamics on graphs -- wave propagation on networks of plates},
  J.~Phys.~A \textbf{51}, 445101 (2018).
\bibitem{schrader}
  V.~Kostrykin and R.~Schrader, \textit{Kirchhoff's Rule for Quantum Wires},
  J.~Phys.~A \textbf{32}, 595 (1999).
\bibitem{kuchment}
  P.~Kuchment, \textit{Quantum Graphs I. Some basic structures},
  Waves in Random media \textbf{14}, S107 (2004).
\bibitem{smilansky_model}
  U. Smilansky, \textit{Irreversible quantum graphs},
  Waves in Random Media \textbf{14}, S143 – S153 (2004);
  M. Solomyak, \textit{On a differential operator appearing in the theory
    of irreversible quantum
    graphs}, Waves in Random Media \textbf{14}, S173-S185 (2004);
  U.~Smilansky, M.~Solomyak,
  \textit{The quantum graph as a limit of a network of physical wires},
  Contemporary Mathematics \textbf{415}, 283-292 (2006).
\bibitem{bolte}
  J.~Bolte, S.~Endres, 
  \textit{The Trace Formula for Quantum Graphs with General Self Adjoint Boundary Conditions},
  Ann.~H.~Poincar{\'e} \textbf{10}, 189 (2009).
\end{thebibliography}
\end{document}